\documentclass[geosciences,article,accept,oneauthors,pdftex,10pt,a4paper]{Definitions/mdpi} 
\firstpage{1} 
\pubvolume{8}
\issuenum{00}
\articlenumber{310}
\pubyear{2018}
\copyrightyear{2018}
\history{Received: 21 June 2018; Accepted: 18 August 2018; Published: 21 August 2018}
\graphicspath{{./Definitions/}}

\Title{Selective Aggregation Experiments on Planetesimal Formation and Mercury-Like Planets}

\Author{{Gerhard Wurm}}%\orcidA{}}

\AuthorNames{Gerhard Wurm}

\address[1]{%
{University of Duisburg-Essen}, Lotharstr. 1, D-47057 Duisburg, Germany; gerhard.wurm@uni-due.de; \mbox{Tel.: +49-203-379-1641}}%Please provide the department information.

\corres{\hangafter=1 \hangindent=1.05em \hspace{-0.82em}Correspondence: gerhard.wurm@uni-due.de; Tel.: +49-203-379-1641}

\abstract{Much of a planet's composition could be determined right at the onset of formation.  
Laboratory experiments can constrain these early steps. This includes static tensile strength measurements or collisions carried out under Earth's gravity and on various microgravity platforms.
Among the variety of extrasolar planets which eventually form are (Exo)-Mercury, terrestrial planets with high density. If they form in inner protoplanetary disks, high temperature experiments are mandatory but they are still rare. Beyond the initial process of hit-and-stick collisions, some additional selective processing might be needed to explain Mercury.  In analogy to icy worlds, such~planets might, e.g., form in environments which are enriched in iron. This requires methods to separate iron and silicate at early stages.  Photophoresis might be one viable way. Mercury and {Mercury-like planets} %Please confirm intended meaning is retained. o.k.
might also form due to the ferromagnetic properties of iron and mechanisms like magnetic aggregation in disk magnetic fields might become important. This review highlights some of the mechanisms with the potential to trigger Mercury formation.}

\keyword{\textls[-5]{mercury; planet formation; protoplanetary disk; iron-silicate separation; extrasolar planets}}

\begin{document}

\section{Planet Formation}

The problem of planet formation has not been solved. There have been many reviews and books on planet formation over previous years though to highlight the state-of-the-art knowns and unknowns~\citep{Blum2018, Wyatt2018, Kley2017,  Morbidelli2016, Pfalzner2015, Johansen2014,  Armitage2013, Blum2008, Leinhardt2008}. Some of these have focused more on theoretical aspects or numerical modeling. A few have approached the problem from the meteoritical side. Also, some have focused on earlier times of protoplanetary disks while others focused more on the later stages of planetary system evolution. 

{This review is also biased. Here, we highlight processes that would be in favor of the formation of a planet like Mercury. In particular, we put the focus on laboratory experiments simulating the early phases of planet formation. The rough structure of the review is as follows. First, a sketch of what is generally known about planet formation is laid out as the fundamental basis. Then, a laboratory technique is introduced which, in previous years, has proved to be highly useful for studying the interaction of many particle systems of dust grains and aggregates thereof. This might be very specific, but it is not widely known. In particular, the underlying physics that it is based upon is itself important in protoplanetary disks of low pressure. This becomes important again later on. These experiments lead to the bouncing barrier and ways how it might be overcome. This is then the entry point for introducing the special nature of Mercury-like planets. Naturally, the influence of temperature is discussed next, as Mercury is the innermost planet of the solar system that is closest to the sun. To explain the {metal-rich planet} 
photophoresis and thermoluminescent erosion are introduced as metal-selective. Finally, the effects of magnetic selection due to the ferromagnetic nature of iron are discussed based on novel experiments, and the review is closed with a short conclusion}.

\subsection{The Red Lines of Planet Formation}

\begin{figure}[H]
\centering
\includegraphics[width=15 cm]{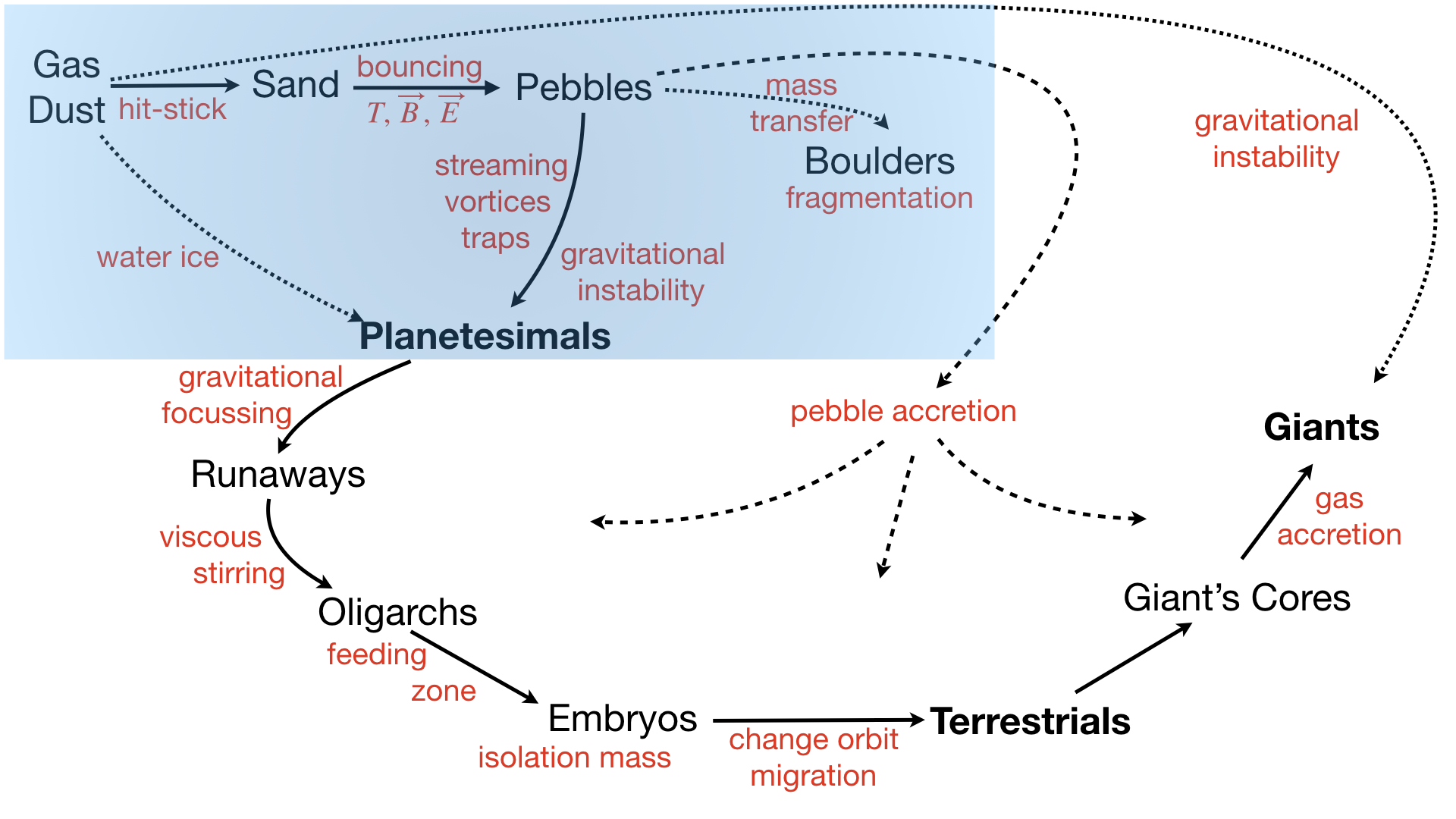}
\caption{\label{planetformation} \textls[-5]{Some standard ideas on planet formation. {The top left, blue shaded region concerns the processing of small grains and is, in part, accessible to laboratory experiments.} Sand, pebble, and~boulder are used for particles of different size ranges. Solid lines indicate standard ways to proceed in size. Dotted lines are alternatives or still under discussion. Dashed ways refer to the influence of pebbles (cm-sized objects), currently studied for a number of different phases of planet formation. Red:~some dominating processes in certain phases; bold: most important milestones}.}
\end{figure}

{\citet{Johansen2014} called their review ``multifaceted planetesimal formation''. Indeed planets and planetesimal formation have many aspects. Nevertheless, some milestones and processes from micrometer dust to planet size objects are accepted by most in the field. Previous research has focused on different details along these few red lines in planet formation.}  Aspects of planet formation {along these lines} that are currently considered to have a reasonable share in planet formation are shown in Figure \ref{planetformation}.

Starting in a protoplanetary disk, gas giants have to form at least within a few million years before the disk is dispersed~\citep{Richert2018, Haisch2001}. In the core accretion scenario, a core that would count as a nice Super Earth by itself has to be formed first to efficiently accrete gas afterwards~\citep{Pollack1996, Alibert2017}. Much of the planet formation in this scenario and for terrestrial planets anyway is a bottom up perspective where dust is the starting point and ever larger bodies evolve one way or the other. The details of ``one way or the other'' is actually the important and interesting, debatable part of planet formation. Laboratory experiments are one possible way to constrain these details, especially, the formation of km-size planetesimals.

{The two mechanisms that have been discussed for planetesimal formation are collisional growth~\citep{Weidenschilling1993} and self-gravity~\citep{Goldreich1973}}.  Initially, all relevant experiments and simulations show that conquering the first few orders of magnitude in size can easily proceed by collisional aggregation. Starting from Brownian motion-induced collisions between sub-micrometer dust grains~\citep{Blum2000a,Blum1996} over fractals \cite{Paszun2006, Wurm1998} to compact sub-mm aggregates \cite{Paszun2009, Weidling2009, Blum2000}, there is no doubt that collisional growth is efficient. Exact timescales are set by the local dust densities and variations in collision velocities related to the disk's gas. If the aggregates are in the sub-mm or mm range after this initial dust merging might be debated, but the first of several hard stops in growth is then encountered. 

\begin{figure}[H]
\centering
\includegraphics[width=12 cm]{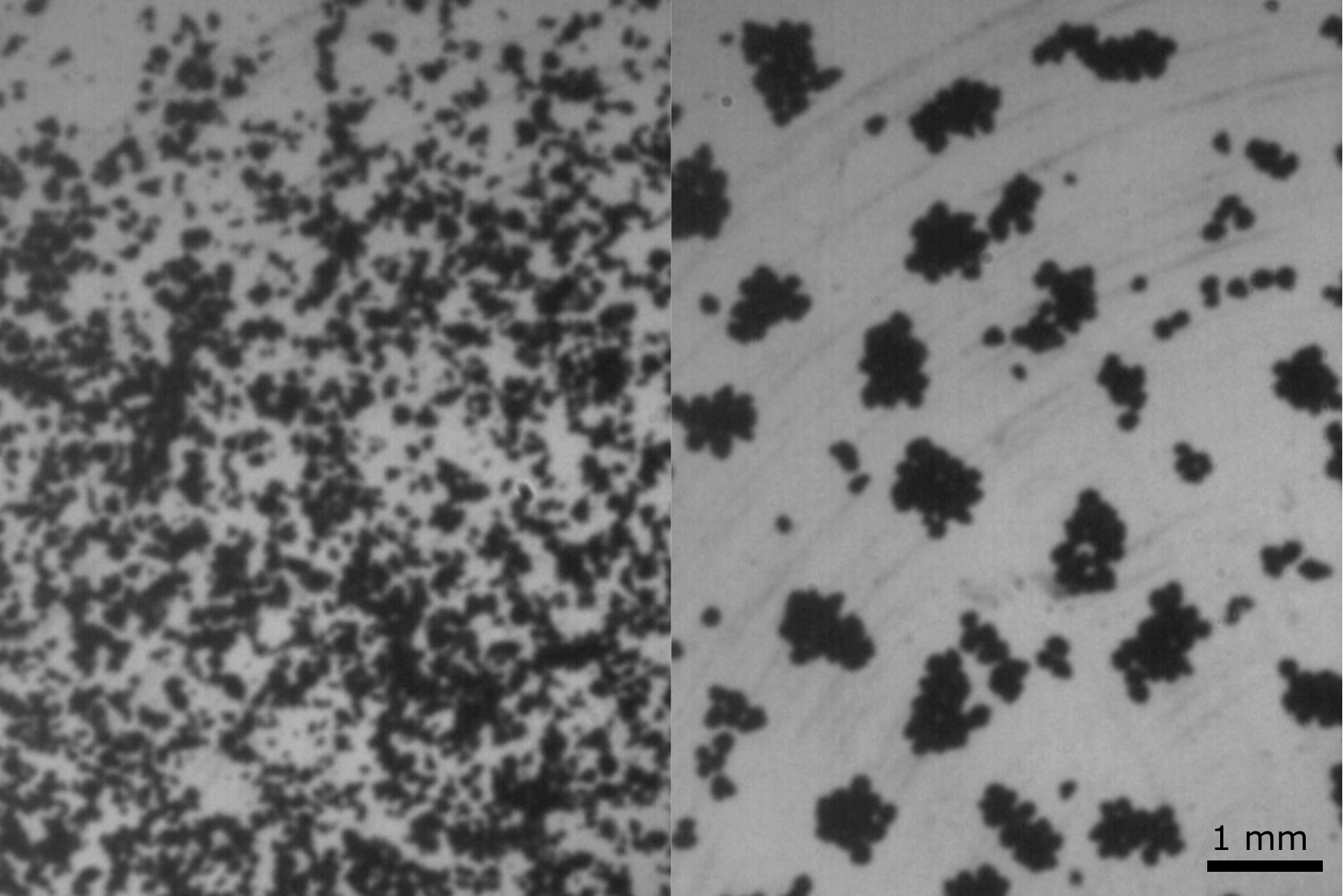}
\caption{\label{bouncing}Initial and final ensemble in a laboratory experiment where dust aggregates are levitated at low ambient pressure on a hot surface by a self-generated Knudsen compressor. Starting with 100 micrometer dust
aggregates, mm-aggregatges grow at speeds of mm/s to cm/s. (\mbox{from~\citet{Demirci2017}}).}
\end{figure}

Once the dust aggregates are compact, the collision energy can essentially no longer be dissipated by restructuring, and collisions lead to bouncing, hence, this was termed the bouncing barrier~\citep{Zsom2010}. Cases~of sticking in individual collision experiments have been observed aplenty ~\citep{Whizin2017, Brisset2017, Jankowski2012}. This~might be taken as a hint that the bouncing barrier might not be so rigid. However, experiments with ensembles of grains showed that such connections between grains are only temporary~\citep{Jankowski2012}. Being very weak, the connections act like rated breakpoints only to be broken by any later collision. In an ever colliding ensemble, one collision can easily disassemble a connected cluster that has formed previously, returning Sisyphos' stone back downhill. It is important to note that in a disk dominated in mass by the gas, the bouncing of solids does not lead to a collisional cooling. The mm grains couple to the gas on timescales far smaller than collision timescales, and collision velocities stay constantly ``high''.
Self-consistent growth experiments starting with smaller grains have shown that a bouncing barrier naturally evolves~\citep{Demirci2017, Kruss2017, Kruss2016, Kelling2014}. A snapshot of an initial and  final configuration from a levitation experiment by \mbox{\citet{Demirci2017}} is shown in Figure \ref{bouncing}.

Relevant collision velocities in protoplanetary disks are only in the mm/s to cm/s range for mm grains. Being slow makes it hard to study free collisions in laboratory experiments. Therefore, two main roads were taken to study such processes. First, microgravity allows slow collisions to be studied without the interference of sedimentation~\citep{Yoshimatsu2017, Kothe2013}. Early on, it was shown that bouncing collisions can prevail among the different outcomes of collisions~\citep{Guettler2010}. The microgravity time that can be used more regularly in facilities, like drop towers or parabolic flights, is short---in the order of, e.g., 9 s for a launched experiment in the Bremen drop tower or somewhat larger than 20 s in parabolic flights. Long term evolution cannot be studied here. This requires sub-orbital flights with microgravity times of minutes~\citep{Brisset2016, Blum1999} or experiments in orbit~\citep{Musiolik2018, Brisset2017, Blum2000a}. Such opportunities are rare though and require long development times.

\section{Experiments with Levitated Grains}

\begin{figure}[H]
\centering
\includegraphics[width=12 cm]{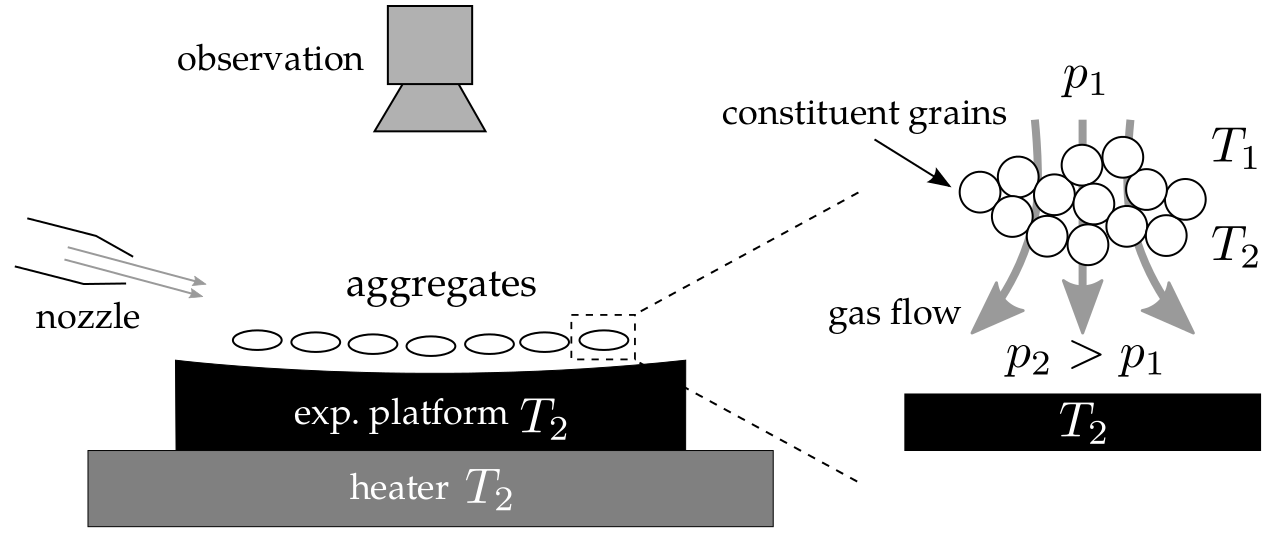}
\caption{\label{knudsen} At low ambient pressures, thermal creep through dust aggregates generates an air cushion for the aggregates to levitate upon (from \citet{Kelling2014}).}
\end{figure}

Alternatively, gravity can be accounted for by different levitation mechanisms. One that was used frequently for studies related to planet formation was introduced by \citet{Kelling2009}. 
{One~might ask if it might be useful to elaborate this specific experimental technique in more detail here. Considering that this is a review, we still think this is valuable, as the underlying physics is related to the later fundamental process of photophoresis or thermoluminescent erosion. It is prime evidence that the effects of low pressure non-equilibrium physics are more than simply a detail as they even allow the Earth's gravity to be counteracted.} 
The levitation principle is therefore visualized in Figure \ref{knudsen}.

The way this works is as follows. Along a surface with a temperature gradient, gas always flows from cold to warm in a thin layer that is 
the size of the mean free path of the molecules. This is called thermal creep. Thermal creep might be slightly counterintuitive. Thinking of an ideal gas being heated, the gas expands, moving from warm to cold. Thermal creep is gas moving along a surface the other way from cold to warm.
Usually, there is a backflow above that layer and thermal creep goes unnoticed at high pressure.
If so, it can safely be ignored. In some cases though, being as remote as it can be for many every-day applications, thermal creep comes with surprises. In a capillary, which is smaller than or on the order of the mean free path, the backflow is suppressed and capillaries can act as pumps, called Knudsen pumps due to some pioneering work by \citet{Knudsen1909}.

The pores of a dust aggregate in the experiments mentioned here act like capillaries. So~if dust aggregates are placed on a heater their bottom heats up. The top sides cool by radiation. If~the pressure is now reduced to some mbar, thermal creep starts working and the dust aggregates become small Knudsen pumps. An overpressure is generated between heater and dust aggregate. If that pressure increase is strong enough---and in fact it can be much stronger than an aggregate's weight---the~aggregate lifts itself. Even ice aggregates can be lifted this way as long as the top side is much cooler e.g., using a liquid nitrogen cooled sky above the ice aggregates though thermophoresis, \mbox{the motion} of a solid particle in a gas with temperature gradient, has to be considered in \mbox{addition~\citep{Kelling2011x, Fung2017, Aumatell2014}.}
\citet{Aumatell2011} used this to study the break-up of sublimating ice aggregates, which is a mechanism discussed to generate dust enrichment close to the snowline~\citep{Sato2011, Schoonenberg2017} or as part of a more elaborate scheme of snowline effects on planet formation~\citep{Drazkowska2017, Morbidelli2016b}.
Anyway, the maximum overpressure in a capillary Knudsen pump can be calculated by~\citep{Muntz2002, DeBeule2014}.

\begin{equation}
\Delta P = P_{avg} \frac{\Delta T}{T_{avg}}\frac{Q_T}{Q_P},
\end{equation}
where $P_{avg}$ and $T_{avg}$ are the average pressure and temperature.
If the aggregate could freely pump and would not create a pressure difference, a continuous flow rate density $\dot{M}$ can be generated according to

\begin{equation}
\dot{M} = P_{avg} \frac{F}{\sqrt{2 k_B/ \mu T_{avg}}} \left( \frac{L}{L_{x}} \frac{\Delta T}{T_{avg}} Q_T  \right),
\end{equation}
with $F$ as fraction of the area covered by capillaries.
In both equations  $Q$ are parameters depending on the Knudsen number Kn (ratio between mean free path and particle size). $Q_T$ harbors the physics of thermal creep. $L$ is the diameter of the capillary or pores, $L_{x}$ is the thickness of the dust layer or length of the capillary.  As mentioned above for high pressure (small Kn) $Q_T$ turns to zero. These equations are displayed here to highlight the important parameters. Details, e.g., on calculating $Q_T$ and $Q_P$ can be found in the respective references.  Based on laboratory experiments~\citet{Koester2017} set up a model of the granular Knudsen compressor accurate to a factor of 2 and details can be found there.

Thermal creep and related effects get some more attention in this review for several reasons. It~drives levitation which proved to be a powerful tool for laboratory studies as indicated above. In~contrast to e.g., a flow table at normal pressure to levitate macroscopic objects, collisions can essentially be studied as ballistic due to the low pressure. 
Second, the same mechanism of thermal creep can also provide internal overpressure in experiments to measure tensile strength. It especially provides a way to produce overpressure in very small, very fragile dust aggregates~\citep{Musiolik2017}.
In~short, if~a weakly bound, warm aggregate cools, it cools outside-in. That generates a temperature gradient and thermal creep. Again gas flows toward the warm side, this time the center of the aggregate. This way, it~generates an overpressure within the aggregate, which can overcome tensile strength~\cite{Musiolik2017}.
Combining~both, levitation and overpressure by cooling allows a systematic study of tensile strengths on the Pa level, e.g., important for cometary matter (Demirci, personal communication).
Maybe even more interesting, as further aspect though, beyond its technical application, low pressure/non-equilibrium gas physics is actually a mechanism acting under the real conditions of protoplanetary disks and planet formation itself. Especially Mercury formation might in parts depend on it. 
Before highlighting some of these aspects we would like to return to the bouncing barrier first, though. 

\subsection{Evolution beyond Bouncing}

How can pre-planetary evolution proceed if, due to bouncing, collisional growth no longer works for sand size particles? There are two options which can both kind of get started with the question: What if we just assume for a moment that not sand but larger pebble or boulder size bodies were present in the disk for some unknown reason?
 
\begin{figure}[H]
\centering
\includegraphics[width=12 cm]{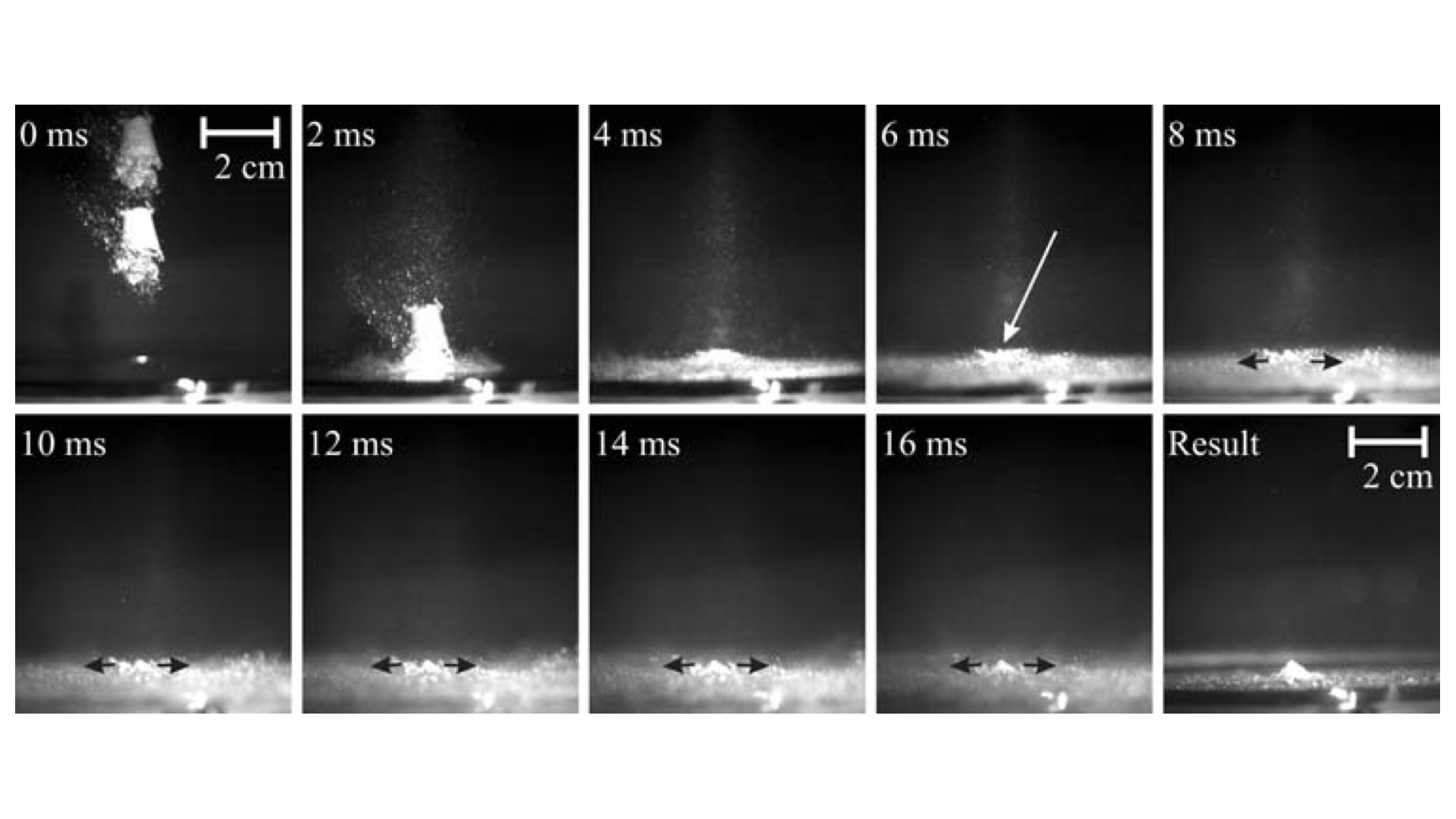}
\vspace{-12pt}
\caption{\label{masstransfer}  If small aggregates collide with larger ones the small fragmenting projectile leaves some mass on the larger target (from  \citet{Teiser2009a}).}
\end{figure}

Also here, as first option, collisions have been considered. The results of collisions change strongly with a larger body present. Larger bodies get more sensitive to turbulence and radial drift velocities increase. Taken together they move with higher velocity through the disk~\citep{Husmann2016, Ormel2007, Weidenschilling1977}. If they collide among themselves they tend to get destroyed easily as has also been seen in experiments~\citep{Deckers2013, Beitz2011, Schraepler2012}. This~fragmentation is generally considered an obstacle for further growth~\citep{Birnstiel2010, Birnstiel2012}. However, if these big aggregates collide with smaller particles, i.e., the sand-sized population stuck at the bouncing barrier, they can still grow~\citep{Wurm2005, Teiser2009a, Beitz2011, Meisner2013}. An example is shown in Figure \ref{masstransfer}.
\vspace{-12pt}

The small collision partner (projectile) does fragment but some part sticks to the larger body (target). This is sometimes called mass transfer collision. This way growth of planetesimals is possible in principle as larger bodies feed upon the smaller grains~\citep{Windmark2012, Drazkowska2014}. It is also interesting that the bouncing barrier is actually beneficial for producing ever larger bodies. If all grains could grow in all collisions there would be many large objects eventually, only to be destroyed again while colliding among themselves. This is prevented if most mass is left within the smaller grains. This process is rather slow though and 100 m bodies would only grow at 3 AU on timescales of a million years ~\citep{Windmark2012}. Work in this field continues. In any case, mass transfer works and it will be an important process in the detailed evolution of solids in protoplanetary disks even if it should turn out not to be the initial process to set off the first wave of planetesimal formation. 

As alternatives, mechanisms have been worked out to concentrate the solids to a level that self-gravity takes over and leads to a direct formation of planetesimal. Most prominent after some initial increase in dust to gas density ``one way or the other'' is the streaming instability~\citep{Youdin2005, Simon2016}. 
As~indicated above, the streaming is also based on the assumption that larger particles would already be present in large quantities, e.g., by drifting inward~\citep{Birnstiel2012}. The streaming instability especially in the inner regions of protoplanetary disks would need grains of cm-size, or ``pebbles'', rather than sub-mm sized ``sand''~\citep{Johansen2014}. There are many active works on the streaming instability. It has to be noted though that they are all of theoretical or numerical nature and are not the focus here. Nevertheless, first attempts to
verify the concept in laboratory experiments are currently underway and indeed streaming like, collective motions in a cloud of grains has been observed~\citep{Schneider2018}.

The connection between collisional and non-collisional formation path is still not very tight if overlapping at all. It is curious that grains so small as sand are already problematic in the making of planets of thousands of km.
So the question of how to proceed after the bouncing barrier is still not settled. What further options are there? 

\section{Mercury-Like Planets}

At this stage, we leave the common ground of planet formation and focus on
Mercury-like planets. In the solar system, there is a general trend for denser terrestrial planets to be found further inward. {For example, regarding the uncompressed density}
, Mercury is very dense at 5.3
$\rm g/cm^3$, Earth~is somewhat less dense at 4.4 $\rm g/cm^3$, and Mars even lighter at \mbox{3.8 $\rm g/cm^3$~\citep{Spohn2001}}. The high density of Mercury is attributed to its large iron core.
In agreement with the radial trend in density attributed to iron, meteorites, originating mostly from the asteroid belt, have much lower iron contents~\citep{Trieloff2006}. There are, by now, a number of extrasolar planets that have a high density, also~indicating that they have large iron cores~\citep{Rappaport2013, Sinukoff2017, Santerne2018, Guenther2017}. How do these high density planets fit into the scheme of planet formation? We do not discuss late stage processing, like giant impacts or evaporation, which would both strip off the mantle of lower density planets \cite{Benz1988, Cameron1985}. Somehow, results from the MESSENGER probe seem incompatible with such very high temperature events~\citep{Peplowski2011}.
Even though the data are still sparse, Mercury-like planets might form far inside of 1 AU, and the process might favour iron.
Therefore, processes that preferentially form these planets might be sensitive to thermal conductivity, assuming iron to be in a metal phase. Growth processes might also be selective for increasing the temperature, or the ferromagnetic nature of iron might be important. We erode these possibilities a bit more in the following sections.

\subsection{Temperature and Collisional Growth}

Material is one way to play around with for planet formation quite generally. It is also well known from experiments that water ice is much stickier than silicates~\citep{Gundlach2015, Aumatell2014, Deckers2016, Musiolik2016}. This led to the idea that beyond the snowline, literally huge, incredibly low density aggregates of nm-size ice grains might grow, circumventing all barriers before they compact~\citep{Kataoka2013, Okuzumi2012x}.
Further out in the disk, where, e.g., $\rm CO_2$ condenses, this advantage of increased stickiness dissolves again as has been shown in collision experiments by \citet{Musiolik2016b}. This changes the evolution of ``dust'' in protoplanetary disks and {its}
appearance~\citep{Pinilla2017}. Nevertheless, a few AU dominated by water ice might have the potential for collisional growth of water-icy planetesimals.

\begin{figure}[H]
\centering
\includegraphics[width=12 cm]{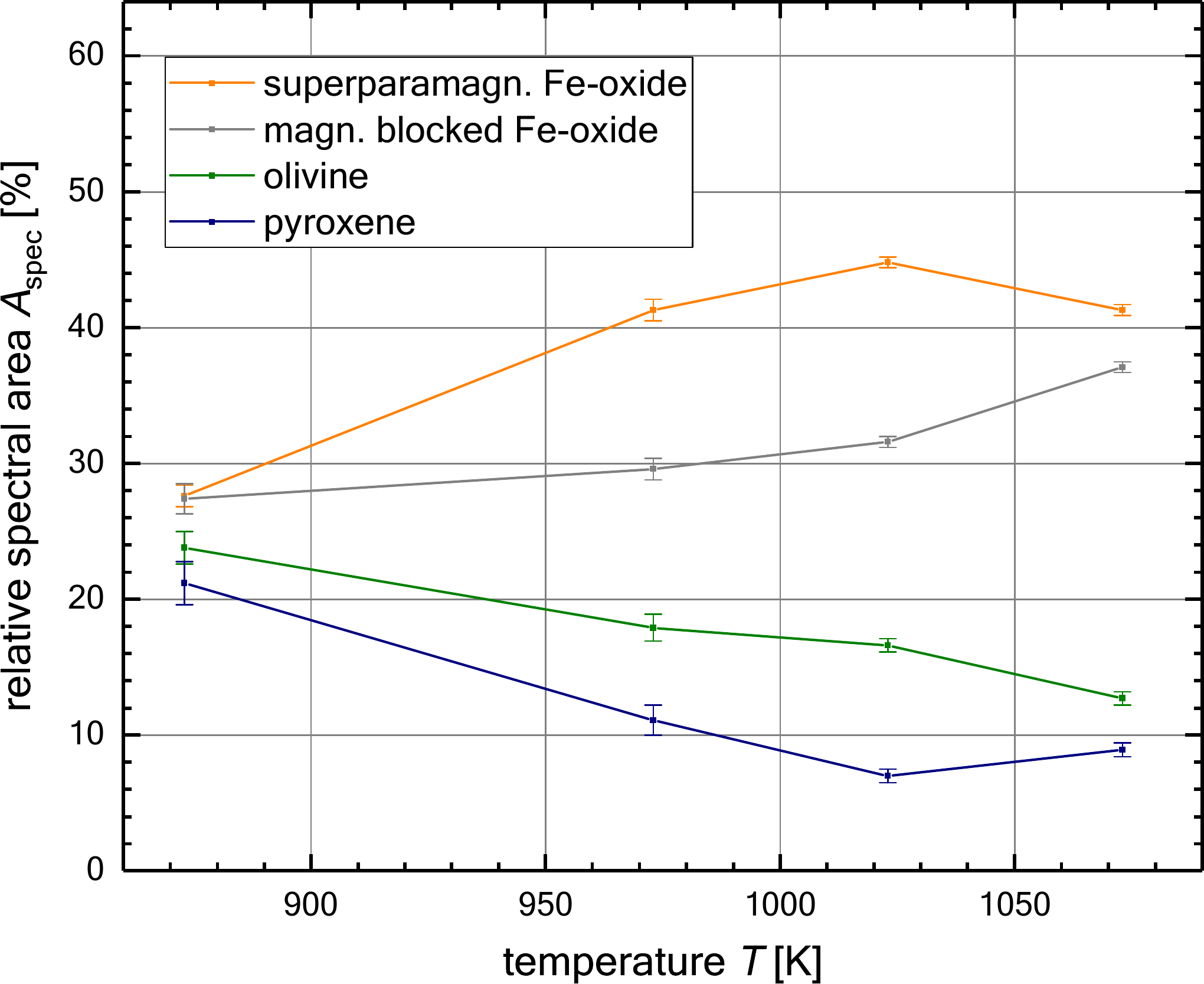}
\caption{\label{mineral} Changes in mineral content measured by Moessbauer spectroscopy for a basalt sample heated in air at different temperatures (from \citet{Demirci2017}).}
\end{figure}

If the inner terrestrial planets, i.e., Mercury, were born locally this cannot be the whole story.
It is still worth following along a similar path and also considering changes in composition in the inner disk, as dust is a multi-component mixture of different minerals. At higher temperatures, mineral transformations are possible, as grains in protoplanetary disks drift inwards towards higher temperatures. 
Radial inward drift of solids naturally occurs in disks with outward decreasing pressure. The~gas in the disk is pressure gradient supported and moves slower than Keplerian. The grains couple to the gas, and they also move slower than Keplerian and thus, feel some residual gravity dragging them inwards~\citep{Weidenschilling1977}. Pressure bumps might locally stop, trap, and concentrate particles in the disk. Anyway, does drift towards higher temperature regions change the composition and does that change sticking?  
{One way to study this in laboratory experiments is as follows: Dust of a given composition that is initially at room temperature is tempered at different higher temperatures for a certain period of time. This~heating might transform minerals. Therefore, the composition of the sample is studied after tempering when the dust is cooled again to room temperature. The same tempered dust samples are used for collision experiments. This way a correlation between sticking properties and radial inward drift in protoplanetary disks can be studied. This approach was taken by \mbox{\citet{Demirci2017}}. They~started with a basalt sample and noted that tempering (in air) induced strong changes in iron bearing minerals as, e.g., seen in Figure \ref{mineral}. They observed a decline in the fractions of olivines and pyroxenes and in the formation of iron oxides.}

\begin{figure}[H]
\centering
\includegraphics[width=12 cm]{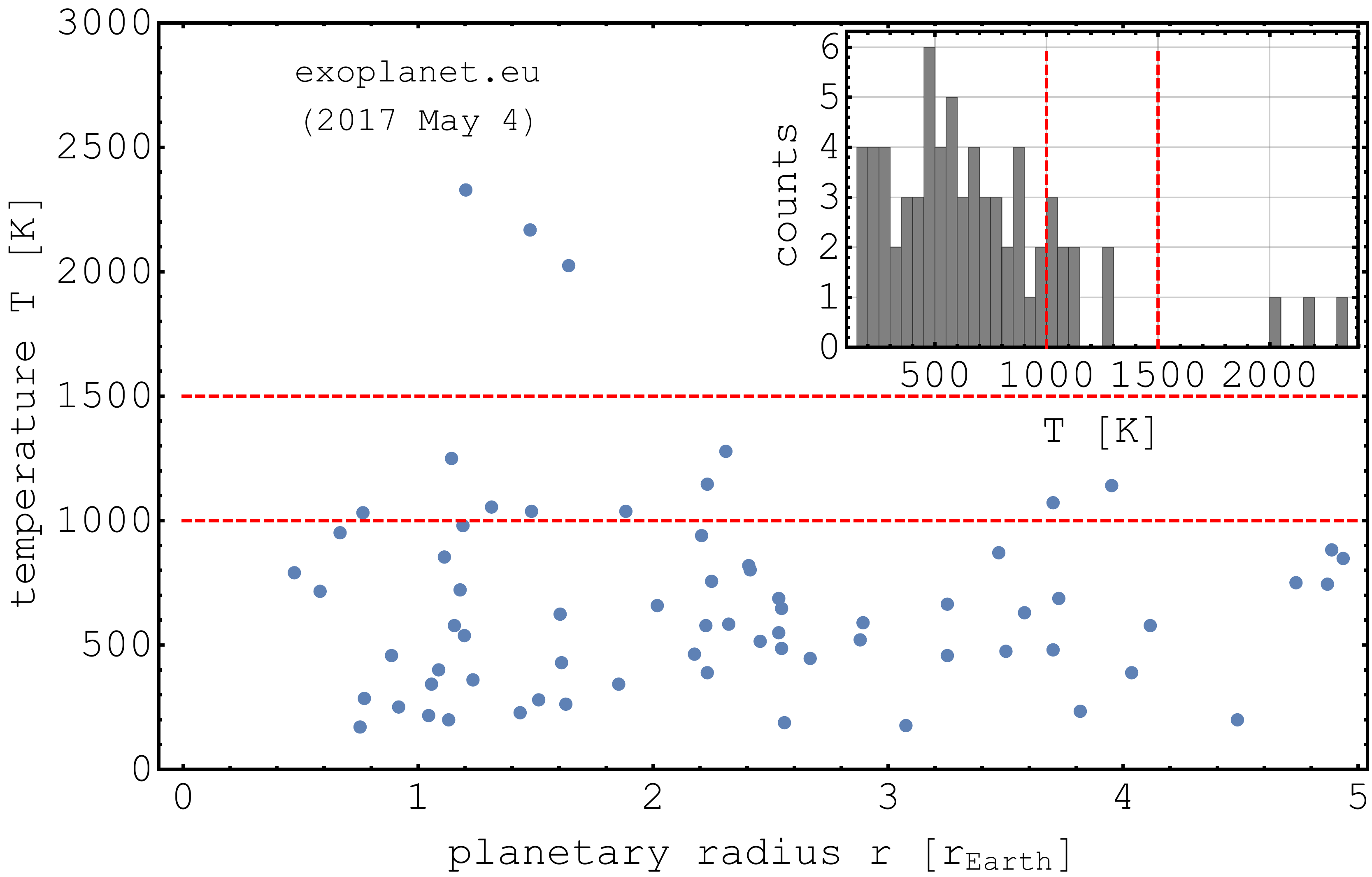}
\caption{\label{thousand} The occurence of small extrasolar planets depending on orbit temperature. The lines mark the sticking transition observed in experiments at 1000 K and the limit of sublimation. {The inset shows the same data as a histogram over temperature} (from \citet{Demirci2017}).}
\end{figure}

{The authors also note that tempering in air does not exactly simulate the solar nebula environment. This will change the mineralogy but, so far, these are the only experiments on this topic, and more relevant ambient environments are planned for the future as the effects on aggregation have been proved to be significant.}
After tempering, collision experiments were carried out. They showed that these mineral changes influence the sticking properties. {While they also noted an increase in the constitutent grain size during tempering which cannot be ruled out completely as a factor, this change was continuous with temperature, while the aggregation changed stepwise.} The size of aggregates grown from dust tempered at more than 1000 K is a factor of 1.5 smaller than for dust not treated at temperatures that high. {Is such a small factor really significant for planetesimal formation? It is, at least, a first hint. On one hand, this is the only experiment that has been carried out so far. It is likely that scanning the large parameter space of minerals, temperatures, ambient gas during tempering, and grain sizes might turn out larger factors. However, even if the factor is that small, not much more might be needed to decide if planetesimals form or not. There is currently only a small gap between the bouncing barrier and streaming instabilities, and small factors might be sufficient.} As seen in Figure \ref{thousand}, exoplanet statistics suggest that they occur less frequently beyond 1000 K. This is still a general statement for planet formation at high temperatures. Whether this favours iron rich minerals in general is an open question.

Real collisions can be complemented by the measurement of static mechanical properties. Tensile~strength, which was mentioned above, is among the important ones. It quantifies a pressure applied to a sample directed outwards to the point that this sample is ripped apart. In a qualitative sense, as it is related to sticking forces, it gives a measure of the capability of a sample to form larger objects in collisions.
They do not give a direct hands-on size-limit for growing dust as directly evident from collision experiments. Nevertheless, tensile strength enters e.g., in collision simulations of larger bodies, which are out of bounds for laboratory experiments~\citep{Schaefer2016, Geretshauser2011, Geretshauser2011a}. 
Systematic measurements were carried out by \citet{Blum2004} for high porosity and \citet{Meisner2012} for untempered, compact $\rm SiO_2$ aggregates.

A first approach to study a higher temperature case was taken by \citet{DeBeule2017}. They found that the tensile strength of tempered palagonite dust increased at about 800 K. This suggests better sticking and growth of larger aggregates, {in contrast to the results found} by \citet{Demirci2017} where the size of the dust aggregated at the bouncing barrier decreased beyond 1000 K. However, the tensile strength might not translate directly to larger aggregate growth, {and they used a different mineral composition. How these static measurements can be interpreted in the context of collisions requires the combined study of static and dynamic properties}. In any case, collision experiments suggest that the formation of extrasolar planets might be decreased inside of the 1000 K distance. While this is still tentative, this is in agreement with extrasolar planet statistics (Figure \ref{thousand}). 

\begin{figure}[H]
\centering
\includegraphics[width=12 cm]{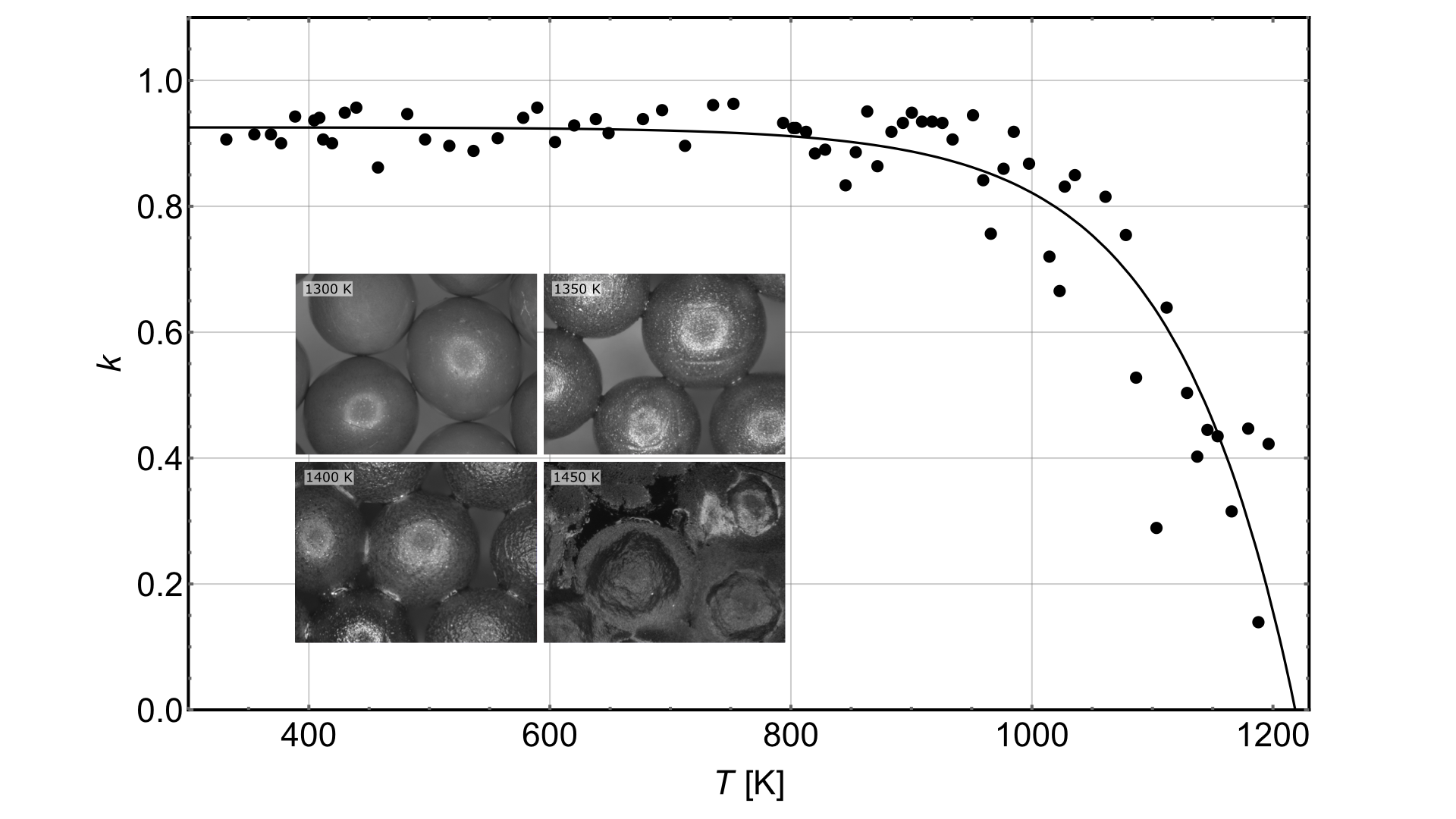}
\caption{\label{cor} Coefficient of restitution ($k$) for collisions between a hot mm basalt sphere and a cold glass target on the ground. The collision velocity was about 1 m/s. The inset shows the onset of sintering and strong melting if heated for 24 h at higher temperatures (from \citet{Bogdan2018}).}
\end{figure}

The experiments mentioned so far simulated mineral transformation, but experiments were carried out under rather cool conditions. So, the dust was only tempered for some time. In addition, effects on collisions and sticking properties have to be expected if the material's viscosity is decreased, or {when the material becomes plastic when grains really collide under hot conditions, i.e., approaching their melting point}. A~very prominent case from the early phases of planet formation showing that two mm particles glue together if they {collide when hot involved compound chondrules}. Chondrules, in general, are formed by flash heating in the solar nebula to temperatures of 1800 K to 1900 K~\citep{Zanda2004}. They are abundant in the meteorite class of chondrites which owe their name to their existence. Chondrules likely were dust aggregates beforehand. Cooling is expected to proceed on a timescale of minutes to hours. 
If~they collide early on during the cooling phase, they collide {in a hot state} 
and form compound chondrules. Early experiments on compound chondrule formation were carried out by \citet{Connolly1994}. Recent collision experiments with mm-glass and basalt grains by \citet{Bogdan2018} approached this from the lower temperature end and with sticking properties in mind. By heating up to 1500 K, they showed that collisions only get stickier above 1000 K, and the coefficient of restitution (velocity after collision/velocity before a collision) only approaches 0 at 1100 K for glass and 1200 K for basalt grains. In any case, no sticking was observed at a collision velocity of about 1 m/s. The coefficient of restitution for basalt is shown in Figure \ref{cor}. Heating at higher temperatures for 24 h allowed grains to sinter together if possible which suggests that the formation of compounds only occurs in a small temperature range between 1350 and 1450 K.

Similar effects have to be expected not only for compound chondrule formation but, in general, in the hot
inner part of protoplanetary disks. Experiments on hot collisions of basalt dust aggregates, once~again using the levitation technique described above but with a different heater (laser), have~already been carried out (Demirci et al., {in preparation}). Up to temperatures of 1100 K, increased sticking probabilities were observed at collisions with speeds in the order of 1~cm/s, though sticking is still an exception. 
{Thus far}
, these experiments have only probed a small range of parameters, and we are still far from a picture of particle growth at high temperatures. In~view of Mercury, a selective view on iron properties during high temperature collisions is needed. 
{It is clearly too early to say whether mineral transformations and growth at high temperatures would favour the formation of {Mercury planets}
. Actually, the few works that have been carried out did not look into this.}
However, it is quite obvious that temperature does not only matter for ice but also at the higher end. If Mercury-like planets form in the inner parts of protoplanetary disks, such effects should be considered for their making, but to what end this leads, remains to be seen. It might support or prohibit planet formation in certain parts of the inner regions of protoplanetary disks.

\subsection{Photophoresis}

Approaching the inner edge of protoplanetary disks, eventually, it has to be considered what effect stellar radiation has on planet formation. It has been shown in a number of papers that photophoresis is an efficient non-gravitational force in such environments, e.g., \citet{Krauss2005} and \citet{Wurm2006} were the first to apply photophoresis to protoplanetary disk conditions. They showed that rings in disks can naturally be formed in optically thin but gaseous disks. Similar aspects were studied by \citet{Herrmann2007}, \citet{Takeuchi2008}, and \citet{Husmann2016} for transitional disks.  When working in transitional disks assumed to be close to dissipation, this might not be a prime mechanism for planet formation,  but rather, {it may be important} 
for shaping the later stages of disks and late stages of planet formation. An important point is that photophoresis requires directed illumination, which, in optical thick parts of protoplanetary disks, {is not provided by the central star}
. In any case,
disks have inner edges in their dust distribution, and these are places where the action of photophoresis is not debatable. 

Photophoresis has some nice features which make it tempting for application to Mercury or iron-rich planet formation. Photophoresis is a force that acts on a particle in a dilute gas due to the temperature difference between the illuminated and non-illuminated sides of a grain \cite{Rohatschek1995, Cordier2016, Beresnev2003zu, Matthews2016}.
This~already hints at its importance in the context of Mercury's formation. As photophoresis depends on the temperature gradient along a grain, it is stronger for particles which do not conduct heat \mbox{well
\citep{Loesche2012}}. Therefore, it clearly distinguishes between metal (iron) grains and silicates. The latter can be pushed outwards by photophoresis while metal grains are left behind. In this way, a natural iron gradient evolves over the protoplanetary disk or initially along its edge, as shown by \mbox{\citet{Wurm2013}}. This compositional differences emerging from photophoretic drag were also seen in more complex simulations of disks~\citep{Cuello2016}. 

While metals are left behind, silicates are transported outwards. This might explain how minerals formed {in a hot state} 
are delivered to the cool comet forming regions~\citep{Krauss2007, Mousis2007, Moudens2011}. If an inner edge is moving outward, photophoresis might trigger a wave of planetesimal formation that is relevant to the meteoritical record~\citep{Haack2007}.
The problem with the optical thick inner part of protoplanetary disks might be bypassed by moving grains over the optical thin surface~\citep{Wurm2009, McNally2017}. 
However, direct stellar radiation might not be the sole driver for photophoresis, e.g., \citet{Loesche2016} showed that temperature fluctuations, e.g., those witnessed by chondrule formation, lead to efficient photophoretic forces. \citet{McNally2015} {investigated photophoresis in more detail in dense environments}
. Therefore, e.g., relative velocities might just be increased for silicates, not for metal grains, giving pure iron a head start in growth.

In any case, focusing on experiments here, besides the numerous numerical simulations that eroded the idea of photophoretic drag, experiments were carried out to quantify the strength for relevant particles in protoplanetary disks.
\citet{Wurm2010} studied the strength for chondrules and dust aggregates in drop tower experiments. \citet{Loesche2014} improved these measurements, \mbox{\citet{Kuepper2014}} quantified photophoresis on very small dust aggregates, and \mbox{\citet{vanBorstel2012}} also added microgravity data about photophoresis. 

\begin{figure}[H]
\centering
\includegraphics[width=12 cm]{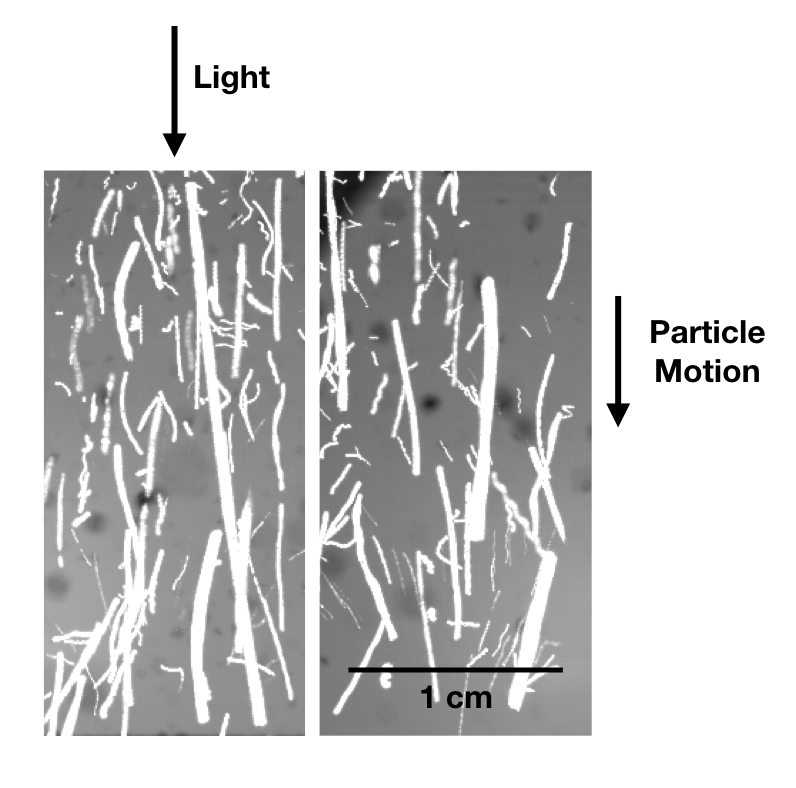}
\vspace{-12pt}
\caption{\label{photo} Photophoresis on JSC Mars 1a grains in a drop tower experiment. Two images with superpositions of 200 frames each from a movie sequence (duration = 0.4 s). The illumination is from the top with a laser beam with 20 kW/$\rm m^2$. The particle velocities depend on the size (smaller grains are generally slower) which is visible in the different track lengths. Larger grains still speed up though, as the gas--grain coupling time is in the order of the elapsed experiment time (see \mbox{\citet{Kuepper2014}} for details of the setup).}
\end{figure}

An example of particle movement in a drop tower experiment is shown in Figure \ref{photo}. 
Grains can easily speed up to 
cm/s, but it should be noted that the large grains still accelerate due to long gas-grain coupling times. {Only once the drag force equals the photophoretic force do particles move with constant velocities.} {At low pressures, the photophoretic strength decreases, but the drag force also decreases}
. Therefore, the final velocities are independent of the ambient gas pressure, and final velocities can be taken as a direct measurement of the expected speeds in protoplanetary disks depending on the light flux. 

While microgravity is well suited to quantify small photophoretic forces, thermal gradient forces can also be stronger than the Earth's gravity~\citep{Kelling2011x, Beresnev2003b}. Therefore, \citet{vanEymeren2012} analyzed the rotation of grains in the context of photophoresis for levitated grains. The result was that rotation is not usually an obstacle to photophoretic forces.

The experiments and numerical simulations led to better quantification of photophoretic forces, as 
given by \citet{Loesche2016a} and \citet{Loesche2016c}, 
i.e., at low protoplanetary disk pressures, photophoretic motion due to temperature differences induced by illumination with a given light flux ($I$) was approximated with high accuracy by~\citep{Loesche2016a}:
\begin{eqnarray}
F = -\frac{\pi}{3}\alpha \frac{P}{T^+ \cdot T}r^2\frac{I J_1}{\frac{k}{r}+4\sigma \epsilon {T_b}^3}\\
T^+ = T+\alpha \left( T_b -T \right)\\
T_b = \sqrt[4]{\frac{I}{4\sigma}+{T}^4}.
\end{eqnarray}

Here, $T$ is the gas (and radiation) temperature; $J_1$ is the asymmetry factor, often assumed to be $J_1 = 0.5$; $\alpha$ is the thermal accomodation factor which is smaller than 1; $\epsilon$ is the emissivity of the grain which is also smaller than 1; $\sigma$ is the Stefan--Boltzmann constant; and $r$ is the particle radius. Importantly in the context of sorting, e.g., due to metal and silicates, is the thermal conductivity ($k$) which can vary by orders of magnitude.

No doubt, in dark disks in thermal equilibrium, photophoresis does not work. However, in the illuminated regions or regions out of equilibrium, photophoresis is not a detail but can, e.g., be stronger than stellar gravity.
We note that photophoresis has also been discussed in atmospheric science~\citep{Beresnev2003b, Cheremisin2005}. Here, a different photophoretic mechanism is often invoked based on differences in the accomodation coefficients over the surface of one individual grain. Then, it is {sufficient} 
for the grain to have a temperature different to the gas temperature, but the grains would move in arbitrary directions or the movement would depend on aligning torques~\citep{Cheremisin2011}. Nevertheless, this might increase the collision velocities and change the early phases of planet formation as well. Experimental findings on levitated grains have sometimes been interpreted as $\alpha$-photophoresis~\citep{Rohatschek1956b, Rohatschek1985Exp, Rohatschek1989}. Specific experiments for dust in protoplanetary environments concerning this $\alpha$-mechanism are missing or have been non-conclusive.

\subsection{Thermoluminescent Erosion of Pre-Planetary Surfaces}

\begin{figure}[H]
\centering
\includegraphics[width=12 cm]{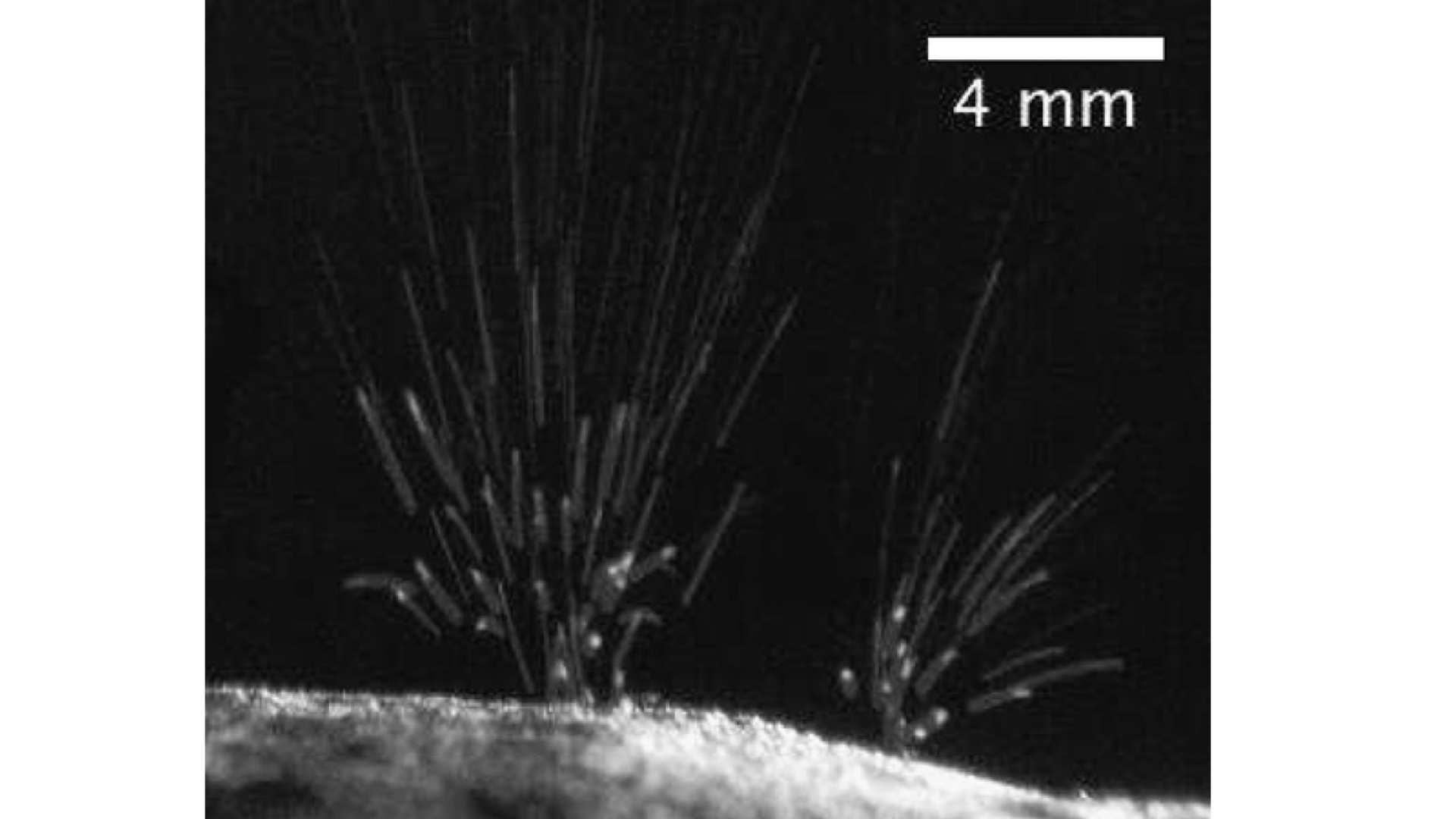}
\caption{\label{undweg} Grains that were eruptively ejected from an illuminated surface at a low ambient pressure (from \citet{Kelling2011}).}
\end{figure}

In the context of inner planets close to the star, light-induced forces are important, as illumination not only pushes grains outwards. A subsurface overpressure can actually destroy larger dusty bodies by continuously shedding its surface layers as experiments and calculations, as shown by \citet{Wurm2007x} and \citet{Wurm2006a}. A number of further experiments and detailed works have been carried out in that field~\citep{Beule2013, Kelling2011, Kocifaj2010, Beule2013}. One of them has already been mentioned above in the context of tensile strength measurements of tempered minerals~\citep{DeBeule2017}.
An example of a particle ejection from a dusty surface at a low ambient pressure is shown in Figure 
\ref{undweg}.

Both, ground-based as well as microgravity experiments have been used to quantify the idea of surface erosion by thermal creep, photophoresis, or thermophoresis. 
In a certain way, this is comparable to the Knudsen compressor levitation experiments described above. Here though, the temperature is set by the illumination. This effect has also been applied to Mars~\citep{Schmidt2017, Debeule2015x, Neakrase2016}. This shows that even moderate illumination of a few hundred $\rm W/m^2$ provides efficient lift. 

In the context of Mercury planets, this mechanism is again sensitive to the material, e.g., heat~conduction. At this point, we can only speculate that, again, selective erosion might leave iron rich planetesimals behind, but experiments on selective erosion are missing.

\subsection{Magnetic Aggregation} 
 
Iron, as a metal, is not only a good heat conductor but also bears the well known 
ferromagnetism. Could this be important for the formation of iron rich planets? Meteorites show that metallic iron was present during the early phases of planet formation. The selective growth of Mercury-like planets might, therefore, be tied to magnetism.
{The underlying process is rather simple. Magnetic dipoles attract each other. This force can be much larger than van-der-Waals forces for sticking grains and {can act over} 
larger distances. Aggregates of magnetic dipoles should, therefore, be formed more easily and should be more stable. If this bias of magnetic aggregation translates into larger aggregate sizes of iron rich material, further processes, e.g., streaming will carry this bias to planetesimal size.}
There are a few works and experiments that have investigated the aggregation of iron~\citep{Nuth1994, Nuebold2003, Dominik2002}. They studied the aggregation of dust without an external magnetic field though, which limits the magnetization to some remanent magnetic fields. Iron can do more. In the inner regions of protoplanetary disks magnetic fields in the order of 1 mT or more can be present~\citep{Dudorov2014, Brauer2017}. This is sufficient to magnetize iron strongly and might lead to the preferential growth of iron aggregates where silicate aggregates would no longer grow. \citet{Hubbard2014} suggested a mechanism that he called magnetic erosion, where the speed up of approaching magnetized dust destroys an aggregate and removes the silicates but keeps the iron bound. 

\begin{figure}[H]
\centering
\includegraphics[width=8.1 cm]{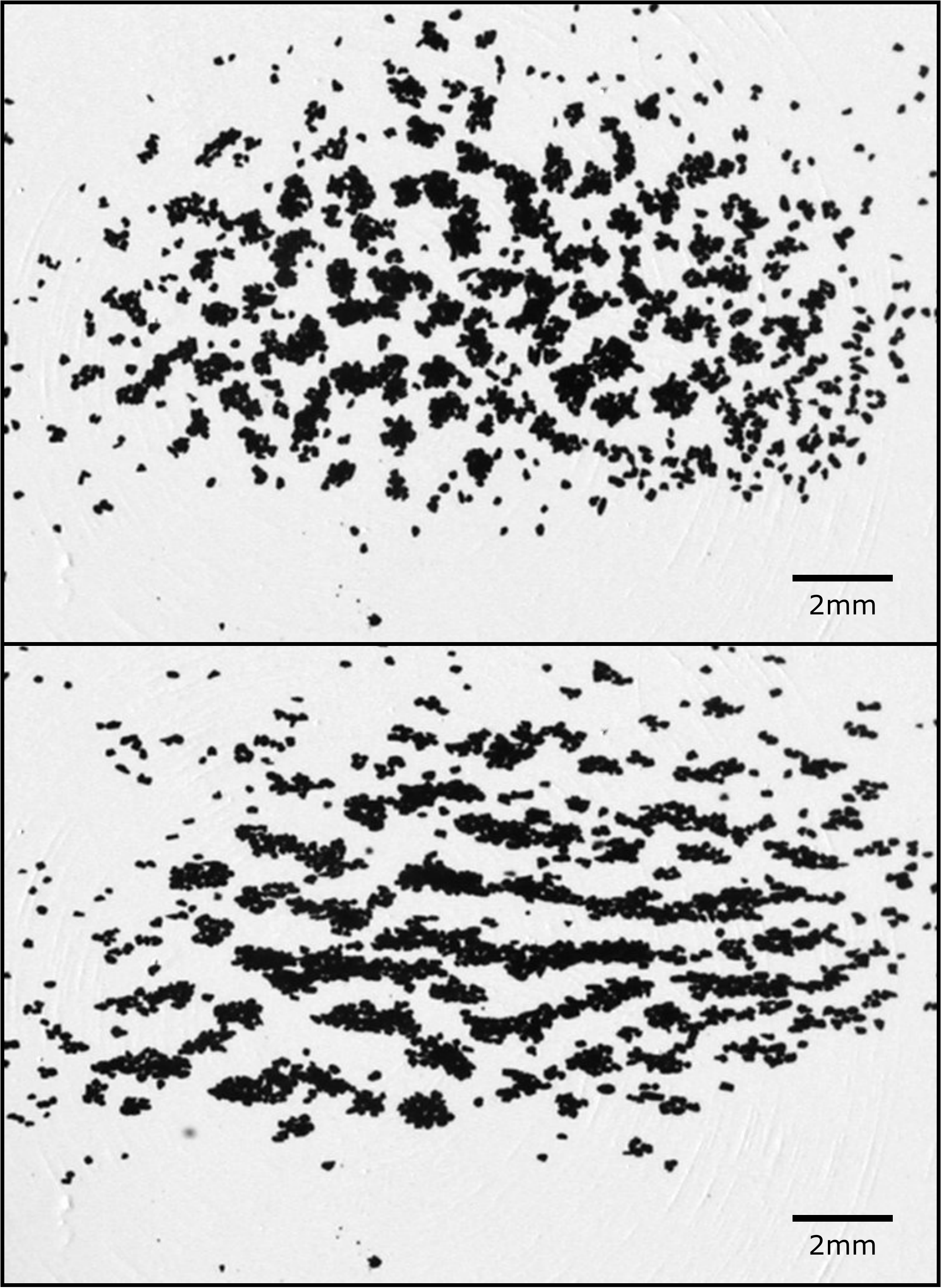}
\caption{\label{magnets} Formation of an elongated cluster of aggregates after applying a magnetic field of 7 mT (from \citet{Kruss2018}).}
\end{figure}

\citet{Kruss2018} just recently carried out first collision experiments with levitated aggregates consisting of a mix of iron and silicates placed in a magnetic field. While magnetic erosion has not been verified so far, another interesting observation was made. Depending on the field strength and the iron content, the aggregates clustered together, forming larger clusters of aggregates as compounds, bound together by magnetic dipole interaction. Turning off the magnetic field, the clusters dissolved again. The clusters of aggregates in a magnetic field turned on and off allows the clustering to be visualized nicely (Figure \ref{magnets}).
 
This mechanism can clearly shift the bouncing barrier to larger sizes, preferentially for metal (iron) rich aggregates. Therefore, below the Curie temperature of 1040 K where iron is ferromagnetic and decreasing with distance to the star, the preferential growth of Mercury-like (iron rich) planets might be initiated. This is also consistent with the general trend of iron depletion with the radial distance. It is also interesting that this mechanism is again tied to a temperature threshold of about 1000 K as also found in the collision experiments with tempered dust and---more important---in the (preliminary) statistics of terrestrial extrasolar planets~\citep{Demirci2017}.
The study of magnetic aggregation has just started, but the first results show promising potential, especially for Mercury formation.

\section{Conclusions}

Laboratory experiments are a valuable tool for testing and quantifying different mechanisms that might be important for planet formation. With a focus on Mercury, processes that act better at small distances to the star and that distinguish between iron and silicates are of special interest. Increasing temperatures are one aspect, and clearly, collisions depend on the temperature---its history as well as the actual temperature during collisions. No concluding remark can be given yet, except that growth changes. More work is needed in this field beyond the handful of experiments carried out.

For the application of photophoresis, light, or at least temperature fluctuation, is a prerequisite. So, in a sense, this mechanism always has to fight for light.  However, if any directed or fluctuating radiation field is present, at least ranging from visible to IR, its effect is huge, and it cannot be neglected. It has the advantage of combining two aspects: it is, in principle, stronger closer to the star and it can separate iron from silicates, both supporting the radial metal gradient and Mercury's iron nature. 

Last but not least, aggregation in magnetic fields also fulfills both requirements. Magnetic fields are stronger closer to the star, and aggregation is favoured for metallic iron dominated grains (below~1040 K).

Laboratory experiments are limited to small size scales. So, these mechanisms can only trigger formation processes. How this is inherited by proceeding formation phases is a different story. 
{At~this point, it is unclear if this all regularly leads to the formation of Mercury planets or if Mercury is rather exceptional.}
However, there are some promising candidate mechanisms supported by laboratory experiments to explain the high density of Mercury and the making of Exo-Mercury planets.

\vspace{6pt}

%%%%%%%%%%%%%%%%%%%%%%%%%%%%%%%%%%%%%%%%%%
\funding{``This research received no external funding.'' }

%%%%%%%%%%%%%%%%%%%%%%%%%%%%%%%%%%%%%%%%%%
\acknowledgments{Many thanks to the three unknown reviewers.}

%%%%%%%%%%%%%%%%%%%%%%%%%%%%%%%%%%%%%%%%%%
\conflictsofinterest{{``The authors declare no conflict of interest.''}}  

%%%%%%%%%%%%%%%%%%%%%%%%%%%%%%%%%%%%%%%%%%
%% optional
%\abbreviations{The following abbreviations are used in this manuscript:\\

%\noindent 
%\begin{tabular}{@{}ll}
%MDPI & Multidisciplinary Digital Publishing Institute\\
%DOAJ & Directory of open access journals\\
%TLA & Three letter acronym\\
%LD & linear dichroism
%\end{tabular}}

%=====================================
% References, variant A: internal bibliography
%=====================================
\reftitle{References}
%\externalbibliography{yes}
%\bibliography{references}

\begin{thebibliography}{999}
\providecommand{\natexlab}[1]{#1}

\bibitem[{Blum}(2018)]{Blum2018}
{Blum}, J.
\newblock {Dust Evolution in Protoplanetary Discs and the Formation of
  Planetesimals. What Have We Learned from Laboratory Experiments?}
\newblock {\em Space Sci. Rev.} {\bf 2018}, {\em 214},~52,
\newblock
  doi:{\changeurlcolor{black}\href{https://doi.org/10.1007/s11214-018-0486-5}{\detokenize{10.1007/s11214-018-0486-5}}}.

\bibitem[{Wyatt}(2018)]{Wyatt2018}
{Wyatt}, M.C.
\newblock {Debris Disks: Probing Planet Formation}.
\newblock {\em ArXiv} {\bf 2018},
  \href{http://xxx.lanl.gov/abs/1804.08636}{{arXiv:1804.08636}}.

\bibitem[{Kley}(2017)]{Kley2017}
{Kley}, W.
\newblock {Planet formation and disk-planet interactions}.
\newblock {\em ArXiv} {\bf 2017},
  \href{http://xxx.lanl.gov/abs/1707.07148}{{  arXiv:1707.07148}}.

\bibitem[{Morbidelli} and {Raymond}(2016)]{Morbidelli2016}
{Morbidelli}, A.; {Raymond}, S.N.
\newblock {Challenges in planet formation}.
\newblock {\em J. Geophys. Res. (Planets)} {\bf 2016}, {\em
  121},~1962--1980, 
\newblock
  doi:{\changeurlcolor{black}\href{https://doi.org/10.1002/2016JE005088}{\detokenize{10.1002/2016JE005088}}}.

\bibitem[{Pfalzner} {et~al.}(2015){Pfalzner}, {Davies}, {Gounelle},
  {Johansen}, {M{\"u}nker}, {Lacerda}, {Portegies Zwart}, {Testi}, {Trieloff},
  and {Veras}]{Pfalzner2015}
{Pfalzner}, S.; {Davies}, M.B.; {Gounelle}, M.; {Johansen}, A.; {M{\"u}nker},
  C.; {Lacerda}, P.; {Portegies Zwart}, S.; {Testi}, L.; {Trieloff}, M.;
  {Veras}, D.
\newblock {The formation of the solar system}.
\newblock {\em Phys. Scr.} {\bf 2015}, {\em 90},~068001,
\newblock
  doi:{\changeurlcolor{black}\href{https://doi.org/10.1088/0031-8949/90/6/068001}{\detokenize{10.1088/0031-8949/90/6/068001}}}.

\bibitem[{Johansen} {et~al.}(2014){Johansen}, {Blum}, {Tanaka}, {Ormel},
  {Bizzarro}, and {Rickman}]{Johansen2014}
{Johansen}, A.; {Blum}, J.; {Tanaka}, H.; {Ormel}, C.; {Bizzarro}, M.;
  {Rickman}, H.
\newblock {The Multifaceted Planetesimal Formation Process}.
\newblock In {\em Protostars and Planets VI}; The University of Arizona Press: Tucson, AZ, USA, {2014}; \mbox{pp. 547--570},
\newblock
  doi:{\changeurlcolor{black}\href{https://doi.org/10.2458/azu_uapress_9780816531240-ch024}{\detokenize{10.2458/azu_uapress_9780816531240-ch024}}}.

\bibitem[{Armitage}(2013)]{Armitage2013}
{Armitage}, P.J.
\newblock {\em {Astrophysics of Planet Formation}};  Cambridge University Press: New York, NY, USA, 2013.

\bibitem[{Blum} and {Wurm}(2008)]{Blum2008}
{Blum}, J.; {Wurm}, G.
\newblock {The Growth Mechanisms of Macroscopic Bodies in Protoplanetary
  Disks}.
\newblock {\em Ann.~Rev. Astron.~Astrophys.} {\bf 2008}, {\em
  46},~21--56,
\newblock
  doi:{\changeurlcolor{black}\href{https://doi.org/10.1146/annurev.astro.46.060407.145152}{\detokenize{10.1146/annurev.astro.46.060407.145152}}}.

\bibitem[{Leinhardt}(2008)]{Leinhardt2008}
{Leinhardt}, Z.M.
\newblock {Terrestrial Planet Formation: A Review and Current Directions}.
\newblock  In {\em Extreme Solar Systems, Proceedings of the Astronomical Society of
  the Pacific Conference (ASP), Santorini Island, Greece, 25--29 June 2007}; {Fischer}, D., {Rasio}, F.A., {Thorsett}, S.E., {Wolszczan}, A., Eds.; Astronomical Society of the Pacific: San Francisco, CA, USA, 2008; Volume 398, p. 225.

\bibitem[{Richert} {et~al.}(2018){Richert}, {Getman}, {Feigelson}, {Kuhn},
  {Broos}, {Povich}, {Bate}, and {Garmire}]{Richert2018}
{Richert}, A.J.W.; {Getman}, K.V.; {Feigelson}, E.D.; {Kuhn}, M.A.; {Broos},
  P.S.; {Povich}, M.S.; {Bate}, M.R.; {Garmire}, G.P.
\newblock {Circumstellar Disk Lifetimes In Numerous Galactic Young Stellar
  Clusters}.
\newblock {\em ArXiv} {\bf 2018},
  \href{http://xxx.lanl.gov/abs/1804.05076}{{arXiv:1804.05076}}.

\bibitem[{Haisch} {et~al.}(2001){Haisch}, {Lada}, and {Lada}]{Haisch2001}
{Haisch}, Jr., K.E.; {Lada}, E.A.; {Lada}, C.J.
\newblock {Disk Frequencies and Lifetimes in Young Clusters}.
\newblock {\em Astrophys. J.} {\bf 2001}, {\em 553},~L153--L156,
\newblock
  doi:{\changeurlcolor{black}\href{https://doi.org/10.1086/320685}{\detokenize{10.1086/320685}}}.

\bibitem[{Pollack} {et~al.}(1996){Pollack}, {Hubickyj}, {Bodenheimer},
  {Lissauer}, {Podolak}, and {Greenzweig}]{Pollack1996}
{Pollack}, J.B.; {Hubickyj}, O.; {Bodenheimer}, P.; {Lissauer}, J.J.;
  {Podolak}, M.; {Greenzweig}, Y.
\newblock {Formation of the Giant Planets by Concurrent Accretion of Solids and
  Gas}.
\newblock {\em Icarus} {\bf 1996}, {\em 124},~62--85,
\newblock
  doi:{\changeurlcolor{black}\href{https://doi.org/10.1006/icar.1996.0190}{\detokenize{10.1006/icar.1996.0190}}}.

\bibitem[{Alibert}(2017)]{Alibert2017}
{Alibert}, Y.
\newblock {Maximum mass of planetary embryos that formed in core-accretion
  models}.
\newblock {\em Astron. Astrophys.} {\bf 2017}, {\em 606},~A69,
\newblock
  doi:{\changeurlcolor{black}\href{https://doi.org/10.1051/0004-6361/201630051}{\detokenize{10.1051/0004-6361/201630051}}}.

\bibitem[{Weidenschilling} and {Cuzzi}(1993)]{Weidenschilling1993}
{Weidenschilling}, S.J.; {Cuzzi}, J.N.
\newblock {Formation of Planetesimals in the Solar Nebula}.
\newblock  In {\em Protostars and Planets III}; {Levy}, E.H., {Lunine}, J.I., Eds.;
  The University of Arizona Press: Tucson, AZ, USA, 1993; pp. 1031--1060.

\bibitem[{Goldreich} and {Ward}(1973)]{Goldreich1973}
\textls[-25]{{Goldreich}, P.; {Ward}, W.R.
\newblock {The Formation of Planetesimals}.
\newblock {\em Astrophys. J.} {\bf 1973}, {\em 183},~1051--1062,
\newblock
  doi:{\changeurlcolor{black}\href{https://doi.org/10.1086/152291}{\detokenize{10.1086/152291}}}.}

\bibitem[{Blum} {et~al.}(2000){Blum}, {Wurm}, {Kempf}, {Poppe}, {Klahr},
  {Kozasa}, {Rott}, {Henning}, {Dorschner}, {Schr{\"a}pler}, {Keller},
  {Markiewicz}, {Mann}, {Gustafson}, {Giovane}, {Neuhaus}, {Fechtig},
  {Gr{\"u}n}, {Feuerbacher}, {Kochan}, {Ratke}, {El Goresy}, {Morfill},
  {Weidenschilling}, {Schwehm}, {Metzler}, and {Ip}]{Blum2000a}
{Blum}, J.; {Wurm}, G.; {Kempf}, S.; {Poppe}, T.; {Klahr}, H.; {Kozasa}, T.;
  {Rott}, M.; {Henning}, T.; {Dorschner}, J.; {Schr{\"a}pler},~R.; et al.
\newblock {Growth and Form of Planetary Seedlings: Results from a Microgravity
  Aggregation Experiment}.
\newblock {\em Phys. Rev. Lett.} {\bf 2000}, {\em 85},~2426--2429,
\newblock
  doi:{\changeurlcolor{black}\href{https://doi.org/10.1103/PhysRevLett.85.2426}{\detokenize{10.1103/PhysRevLett.85.2426}}}.

\bibitem[{Blum} {et~al.}(1996){Blum}, {Wurm}, {Kempf}, and
  {Henning}]{Blum1996}
{Blum}, J.; {Wurm}, G.; {Kempf}, S.; {Henning}, T.
\newblock {The Brownian Motion of Dust Particles in the Solar Nebula: An
  Experimental Approach to the Problem of Pre-planetary Dust Aggregation}.
\newblock {\em Icarus} {\bf 1996}, {\em 124},~441--451,
\newblock
  doi:{\changeurlcolor{black}\href{https://doi.org/10.1006/icar.1996.0221}{\detokenize{10.1006/icar.1996.0221}}}.

\bibitem[{Paszun} and {Dominik}(2006)]{Paszun2006}
{Paszun}, D.; {Dominik}, C.
\newblock {The influence of grain rotation on the structure of dust
  aggregates}.
\newblock {\em Icarus} {\bf 2006}, {\em 182},~274--280,
\newblock
  doi:{\changeurlcolor{black}\href{https://doi.org/10.1016/j.Icarus.2005.12.018}{\detokenize{10.1016/j.Icarus.2005.12.018}}}.

\bibitem[{Wurm} and {Blum}(1998)]{Wurm1998}
\textls[-50]{{Wurm}, G.; {Blum}, J.
\newblock {Experiments on Preplanetary Dust Aggregation}.
\newblock {\em Icarus} {\bf 1998}, {\em 132},~125--136,
\newblock
  doi:{\changeurlcolor{black}\href{https://doi.org/10.1006/icar.1998.5891}{\detokenize{10.1006/icar.1998.5891}}}.}

\bibitem[{Paszun} and {Dominik}(2009)]{Paszun2009}
\textls[-5]{{Paszun}, D.; {Dominik}, C.
\newblock {Collisional evolution of dust aggregates. From compaction to
  catastrophic~destruction}.
\newblock {\em Astron. Astrophys.} {\bf 2009}, {\em 507},~1023--1040,
\newblock
  doi:{\changeurlcolor{black}\href{https://doi.org/10.1051/0004-6361/200810682}{\detokenize{10.1051/0004-6361/200810682}}}.}

\bibitem[{Weidling} {et~al.}(2009){Weidling}, {G{\"u}ttler}, {Blum}, and
  {Brauer}]{Weidling2009}
{Weidling}, R.; {G{\"u}ttler}, C.; {Blum}, J.; {Brauer}, F.
\newblock {The Physics of Protoplanetesimal Dust Agglomerates. III. Compaction
  in Multiple Collisions}.
\newblock {\em Astrophys. J.} {\bf 2009}, {\em 696},~2036--2043,
\newblock
  doi:{\changeurlcolor{black}\href{https://doi.org/10.1088/0004-637X/696/2/2036}{\detokenize{10.1088/0004-637X/696/2/2036}}}.

\bibitem[{Blum} and {Wurm}(2000)]{Blum2000}
\textls[-15]{{Blum}, J.; {Wurm}, G.
\newblock {Experiments on Sticking, Restructuring, and Fragmentation of
  Preplanetary Dust~Aggregates}.
\newblock {\em Icarus} {\bf 2000}, {\em 143},~138--146,
\newblock
  doi:{\changeurlcolor{black}\href{https://doi.org/10.1006/icar.1999.6234}{\detokenize{10.1006/icar.1999.6234}}}.
}
\bibitem[{Zsom} {et~al.}(2010){Zsom}, {Ormel}, {G{\"u}ttler}, {Blum}, and
  {Dullemond}]{Zsom2010}
{Zsom}, A.; {Ormel}, C.W.; {G{\"u}ttler}, C.; {Blum}, J.; {Dullemond}, C.P.
\newblock {The outcome of protoplanetary dust growth: Pebbles, boulders, or
  planetesimals? II. Introducing the bouncing barrier}.
\newblock {\em Astron. Astrophys.} {\bf 2010}, {\em 513},~A57,
\newblock
  doi:{\changeurlcolor{black}\href{https://doi.org/10.1051/0004-6361/200912976}{\detokenize{10.1051/0004-6361/200912976}}}.

\bibitem[{Whizin} {et~al.}(2017){Whizin}, {Blum}, and {Colwell}]{Whizin2017}
{Whizin}, A.D.; {Blum}, J.; {Colwell}, J.E.
\newblock {The Physics of Protoplanetesimal Dust Agglomerates. VIII.
  Microgravity Collisions between Porous SiO${_2}$ Aggregates and Loosely Bound
  Agglomerates}.
\newblock {\em Astrophys. J.} {\bf 2017}, {\em 836},~94,
\newblock
  doi:{\changeurlcolor{black}\href{https://doi.org/10.3847/1538-4357/836/1/94}{\detokenize{10.3847/1538-4357/836/1/94}}}.

\bibitem[{Brisset} {et~al.}(2017){Brisset}, {Colwell}, {Dove}, and
  {Maukonen}]{Brisset2017}
{Brisset}, J.; {Colwell}, J.; {Dove}, A.; {Maukonen}, D.
\newblock {NanoRocks: Design and performance of an experiment studying planet
  formation on the International Space Station}.
\newblock {\em Rev. Sci. Instrum.} {\bf 2017}, {\em 88},~074502,
\newblock
  doi:{\changeurlcolor{black}\href{https://doi.org/10.1063/1.4991857}{\detokenize{10.1063/1.4991857}}}.

\bibitem[{Jankowski} {et~al.}(2012){Jankowski}, {Wurm}, {Kelling}, {Teiser},
  {Sabolo}, {Guti{\'e}rrez}, and {Bertini}]{Jankowski2012}
{Jankowski}, T.; {Wurm}, G.; {Kelling}, T.; {Teiser}, J.; {Sabolo}, W.;
  {Guti{\'e}rrez}, P.J.; {Bertini}, I.
\newblock {Crossing barriers in planetesimal formation: The growth of mm-dust
  aggregates with large constituent grains}.
\newblock {\em Astron.~Astrophys.} {\bf 2012}, {\em 542},~A80,
\newblock
  doi:{\changeurlcolor{black}\href{https://doi.org/10.1051/0004-6361/201218984}{\detokenize{10.1051/0004-6361/201218984}}}.

\bibitem[{Demirci} {et~al.}(2017){Demirci}, {Teiser}, {Steinpilz},
  {Landers}, {Salamon}, {Wende}, and {Wurm}]{Demirci2017}
{Demirci}, T.; {Teiser}, J.; {Steinpilz}, T.; {Landers}, J.; {Salamon}, S.;
  {Wende}, H.; {Wurm}, G.
\newblock {Is There a Temperature Limit in Planet Formation at 1000 K?}
\newblock {\em Astrophys. J.} {\bf 2017}, {\em 846},~48,
\newblock
  doi:{\changeurlcolor{black}\href{https://doi.org/10.3847/1538-4357/aa816c}{\detokenize{10.3847/1538-4357/aa816c}}}.

\bibitem[{Kruss} {et~al.}(2017){Kruss}, {Teiser}, and {Wurm}]{Kruss2017}
{Kruss}, M.; {Teiser}, J.; {Wurm}, G.
\newblock {Growing into and out of the bouncing barrier in planetesimal
  formation}.
\newblock {\em Astron.~Astrophys.} {\bf 2017}, {\em 600},~A103,
\newblock
  doi:{\changeurlcolor{black}\href{https://doi.org/10.1051/0004-6361/201630251}{\detokenize{10.1051/0004-6361/201630251}}}.

\bibitem[{Kruss} {et~al.}(2016){Kruss}, {Demirci}, {Koester}, {Kelling}, and
  {Wurm}]{Kruss2016}
{Kruss}, M.; {Demirci}, T.; {Koester}, M.; {Kelling}, T.; {Wurm}, G.
\newblock {Failed Growth at the Bouncing Barrier in Planetesimal Formation}.
\newblock {\em Astrophys. J.} {\bf 2016}, {\em 827},~110,
\newblock
  doi:{\changeurlcolor{black}\href{https://doi.org/10.3847/0004-637X/827/2/110}{\detokenize{10.3847/0004-637X/827/2/110}}}.

\bibitem[{Kelling} {et~al.}(2014){Kelling}, {Wurm}, and
  {K{\"o}ster}]{Kelling2014}
{Kelling}, T.; {Wurm}, G.; {K{\"o}ster}, M.
\newblock {Experimental Study on Bouncing Barriers in Protoplanetary Disks}.
\newblock {\em Astrophys. J.} {\bf 2014}, {\em 783},~111,
\newblock
  doi:{\changeurlcolor{black}\href{https://doi.org/10.1088/0004-637X/783/2/111}{\detokenize{10.1088/0004-637X/783/2/111}}}.

\bibitem[{Yoshimatsu} {et~al.}(2017){Yoshimatsu}, {Ara{\'u}jo}, {Wurm},
  {Herrmann}, and {Shinbrot}]{Yoshimatsu2017}
{Yoshimatsu}, R.; {Ara{\'u}jo}, N.A.M.; {Wurm}, G.; {Herrmann}, H.J.;
  {Shinbrot}, T.
\newblock {Self-charging of identical grains in the absence of an external
  field}.
\newblock {\em Sci. Rep.} {\bf 2017}, {\em 7},~39996,
\newblock
  doi:{\changeurlcolor{black}\href{https://doi.org/10.1038/srep39996}{\detokenize{10.1038/srep39996}}}.

\bibitem[{Kothe} {et~al.}(2013){Kothe}, {Blum}, {Weidling}, and
  {G{\"u}ttler}]{Kothe2013}
\textls[-35]{{Kothe}, S.; {Blum}, J.; {Weidling}, R.; {G{\"u}ttler}, C.
\newblock {Free collisions in a microgravity many-particle experiment. III. The
  collision behavior of sub-millimeter-sized dust aggregates}.
\newblock {\em Icarus} {\bf 2013}, {\em 225},~75--85,
\newblock
  doi:{\changeurlcolor{black}\href{https://doi.org/10.1016/j.Icarus.2013.02.034}{\detokenize{10.1016/j.Icarus.2013.02.034}}}.}

\bibitem[{G{\"u}ttler} {et~al.}(2010){G{\"u}ttler}, {Blum}, {Zsom}, {Ormel},
  and {Dullemond}]{Guettler2010}
{G{\"u}ttler}, C.; {Blum}, J.; {Zsom}, A.; {Ormel}, C.W.; {Dullemond}, C.P.
\newblock {The outcome of protoplanetary dust growth: Pebbles, boulders, or
  planetesimals? I. Mapping the zoo of laboratory collision experiments}.
\newblock {\em Astron.~Astrophys.} {\bf 2010}, {\em 513},~A56,
\newblock
  doi:{\changeurlcolor{black}\href{https://doi.org/10.1051/0004-6361/200912852}{\detokenize{10.1051/0004-6361/200912852}}}.

\bibitem[{Brisset} {et~al.}(2016){Brisset}, {Hei{\ss}elmann}, {Kothe},
  {Weidling}, and {Blum}]{Brisset2016}
{Brisset}, J.; {Hei{\ss}elmann}, D.; {Kothe}, S.; {Weidling}, R.; {Blum}, J.
\newblock {Submillimetre-sized dust aggregate collision and growth properties.
  Experimental study of a multi-particle system on a suborbital rocket}.
\newblock {\em Astron.~Astrophys.} {\bf 2016}, {\em 593},~A3,
\newblock
  doi:{\changeurlcolor{black}\href{https://doi.org/10.1051/0004-6361/201527288}{\detokenize{10.1051/0004-6361/201527288}}}.

\bibitem[{Blum} {et~al.}(1999){Blum}, {Wurm}, and {Poppe}]{Blum1999}
{Blum}, J.; {Wurm}, G.; {Poppe}, T.
\newblock {The CODAG sounding rocket experiment to study aggregation of
  thermally diffusing dust particles}.
\newblock {\em Adv. Space Res,} {\bf 1999}, {\em 23},~1267--1270,
\newblock
  doi:{\changeurlcolor{black}\href{https://doi.org/10.1016/S0273-1177(99)00195-7}{\detokenize{10.1016/S0273-1177(99)00195-7}}}.

\bibitem[{Musiolik} {et~al.}(2018){Musiolik}, {Steinpilz}, M., {Jungmann},
  {Demirci}, {Aderholz}, {Teiser}, and {Wurm}]{Musiolik2018}
{Musiolik}, G.; {Steinpilz}, T.; {Kruss}, M.; {Jungmann}, F.; {Demirci}, T.;
  {Aderholz}, M.; {Teiser}, J.; {Wurm}, G.
\newblock {ARISE:~Building Planetary Seedlings on the ISS}.
\newblock  In proceedings of the IAC Conference, Bremen, Germany, \mbox{1--5 October 2018}.

\bibitem[{Kelling} and {Wurm}(2009)]{Kelling2009}
\textls[-25]{{Kelling}, T.; {Wurm}, G.
\newblock {Self-Sustained Levitation of Dust Aggregate Ensembles by
  Temperature-Gradient-Induced Overpressures}.
\newblock {\em Phys. Rev. Lett.} {\bf 2009}, {\em 103},~215502-1--215502-4,
\newblock
  doi:{\changeurlcolor{black}\href{https://doi.org/10.1103/PhysRevLett.103.215502}{\detokenize{10.1103/PhysRevLett.103.215502}}}.}

\bibitem[{Knudsen}(1909)]{Knudsen1909}
{Knudsen}, M.
\newblock {Thermischer Molekulardruck der Gase in R{\"o}hren and por{\"o}sen
  K{\"o}rpern}.
\newblock {\em Ann. Phys.} {\bf 1909}, {\em 336},~633--640,
\newblock
  doi:{\changeurlcolor{black}\href{https://doi.org/10.1002/andp.19103360310}{\detokenize{10.1002/andp.19103360310}}}.

\bibitem[{Kelling} {et~al.}(2011){Kelling}, {Wurm}, and
  {D{\"u}rmann}]{Kelling2011x}
{{Kelling}}, T.; {Wurm}, G.; {D{\"u}rmann}, C.
\newblock {Ice particles trapped by temperature gradients at mbar pressure}.
\newblock {\em Rev.~Sci. Instrum.} {\bf 2011}, {\em 82},~115105,
\newblock
  doi:{\changeurlcolor{black}\href{https://doi.org/10.1063/1.3658824}{\detokenize{10.1063/1.3658824}}}.%ref 39 and 115 are the same. ... removed 115

\bibitem[{Fung} {et~al.}(2017){Fung}, {Usatyuk}, {DeSalvo}, and
  {Chin}]{Fung2017}
{Fung}, F.; {Usatyuk}, M.; {DeSalvo}, B.J.; {Chin}, C.
\newblock {Stable thermophoretic trapping of generic particles at \mbox{low
  pressures}}.
\newblock {\em Appl. Phys. Lett.} {\bf 2017}, {\em 110},~034102,
\newblock
  doi:{\changeurlcolor{black}\href{https://doi.org/10.1063/1.4974489}{\detokenize{10.1063/1.4974489}}}.

\bibitem[{Aumatell} and {Wurm}(2014)]{Aumatell2014}
{Aumatell}, G.; {Wurm}, G.
\newblock {Ice aggregate contacts at the nm-scale}.
\newblock {\em Mon. Not. R. Astron. Soc.} {\bf 2014},
  {\em 437},~690--702,
\newblock doi:{\changeurlcolor{black}\href{https://doi.org/10.1093/Monthly
  Notices of the Royal Astronomical
  Society/stt1921}{\detokenize{10.1093/mnras/stt1921}}}.

\bibitem[{Aumatell} and {Wurm}(2011)]{Aumatell2011}
\textls[-15]{{Aumatell}, G.; {Wurm}, G.
\newblock {Breaking the ice: Planetesimal formation at the snowline}.
\newblock {\em Mon. Not. R. Astron.~Soc.} {\bf 2011},
  {\em 418},~L1--L5,
\newblock
  doi:{\changeurlcolor{black}\href{https://doi.org/10.1111/j.1745-3933.2011.01126.x}{\detokenize{10.1111/j.1745-3933.2011.01126.x}}}.}

\bibitem[{Saito} and {Sirono}(2011)]{Sato2011}
\textls[-55]{{Saito}, E.; {Sirono}, S.i.
\newblock {Planetesimal Formation by Sublimation}.
\newblock {\em Astrophys. J.} {\bf 2011}, {\em 728},~20,
\newblock
  doi:{\changeurlcolor{black}\href{https://doi.org/10.1088/0004-637X/728/1/20}{\detokenize{10.1088/0004-637X/728/1/20}}}.}

\bibitem[{Schoonenberg} and {Ormel}(2017)]{Schoonenberg2017}
{Schoonenberg}, D.; {Ormel}, C.W.
\newblock {Planetesimal formation near the snowline: In or out?}
\newblock {\em Astron. Astrophys.} {\bf 2017}, {\em 602},~A21,
\newblock
  doi:{\changeurlcolor{black}\href{https://doi.org/10.1051/0004-6361/201630013}{\detokenize{10.1051/0004-6361/201630013}}}.

\bibitem[{Dr{\c a}{\.z}kowska} and {Alibert}(2017)]{Drazkowska2017}
{Dr{\c a}{\.z}kowska}, J.; {Alibert}, Y.
\newblock {Planetesimal formation starts at the snow line}.
\newblock {\em Astron. Astrophys.} {\bf 2017}, {\em 608},~A92,
\newblock
  doi:{\changeurlcolor{black}\href{https://doi.org/10.1051/0004-6361/201731491}{\detokenize{10.1051/0004-6361/201731491}}}.

\bibitem[{Morbidelli} {et~al.}(2016){Morbidelli}, {Bitsch}, {Crida},
  {Gounelle}, {Guillot}, {Jacobson}, {Johansen}, {Lambrechts}, and
  {Lega}]{Morbidelli2016b}
{Morbidelli}, A.; {Bitsch}, B.; {Crida}, A.; {Gounelle}, M.; {Guillot}, T.;
  {Jacobson}, S.; {Johansen}, A.; {Lambrechts}, M.; {Lega}, E.
\newblock {Fossilized condensation lines in the Solar System protoplanetary
  disk}.
\newblock {\em Icarus} {\bf 2016}, {\em 267},~368--376,
\newblock
  doi:{\changeurlcolor{black}\href{https://doi.org/10.1016/j.Icarus.2015.11.027}{\detokenize{10.1016/j.Icarus.2015.11.027}}}.

\bibitem[{Muntz} {et~al.}(2002){Muntz}, {Sone}, {Aoki}, {Vargo}, and
  {Young}]{Muntz2002}
{Muntz}, E.P.; {Sone}, Y.; {Aoki}, K.; {Vargo}, S.; {Young}, M.
\newblock Performance analysis and optimization considerations for a Knudsen
  compressor in transitional flow.
\newblock {\em J. Vacuum Sci. Technol. Vac. Surf. Films} {\bf 2002}, {\em 20},~214--224.

\bibitem[{De Beule} {et~al.}(2014){de Beule}, {Wurm}, {Kelling},
  {K{\"u}pper}, {Jankowski}, and {Teiser}]{DeBeule2014}
{de Beule}, C.; {Wurm}, G.; {Kelling}, T.; {K{\"u}pper}, M.; {Jankowski}, T.;
  {Teiser}, J.
\newblock {The martian soil as a planetary gas~pump}.
\newblock {\em Nat. Phys.} {\bf 2014}, {\em 10},~17--20,
\newblock
  doi:{\changeurlcolor{black}\href{https://doi.org/10.1038/nphys2821}{\detokenize{10.1038/nphys2821}}}.

\bibitem[{Koester} {et~al.}(2017){Koester}, {Kelling}, {Teiser}, and
  {Wurm}]{Koester2017}
{Koester}, M.; {Kelling}, T.; {Teiser}, J.; {Wurm}, G.
\newblock {Gas flow within Martian soil: Experiments on granular Knudsen
  compressors}.
\newblock {\em Astrophys. Space Sci.} {\bf 2017}, {\em 362},~171,
\newblock
  doi:{\changeurlcolor{black}\href{https://doi.org/10.1007/s10509-017-3154-4}{\detokenize{10.1007/s10509-017-3154-4}}}.

\bibitem[{Musiolik} {et~al.}(2017){Musiolik}, {de Beule}, and
  {Wurm}]{Musiolik2017}
{Musiolik}, G.; {de Beule}, C.; {Wurm}, G.
\newblock {Analog Experiments on Tensile Strength of Dusty and Cometary
  Matter}.
\newblock {\em Icarus} {\bf 2017}, {\em 296},~110--116,
\newblock
  doi:{\changeurlcolor{black}\href{https://doi.org/10.1016/j.Icarus.2017.05.009}{\detokenize{10.1016/j.Icarus.2017.05.009}}}.

\bibitem[{Husmann} {et~al.}(2016){Husmann}, {Loesche}, and
  {Wurm}]{Husmann2016}
{Husmann}, T.; {Loesche}, C.; {Wurm}, G.
\newblock {Self-sustained Recycling in the Inner Dust Ring of Pre-transitional
  Disks}.
\newblock {\em Astrophys. J.} {\bf 2016}, {\em 829},~111,
\newblock
  doi:{\changeurlcolor{black}\href{https://doi.org/10.3847/0004-637X/829/2/111}{\detokenize{10.3847/0004-637X/829/2/111}}}.

\bibitem[{Ormel} and {Cuzzi}(2007)]{Ormel2007}
{Ormel}, C.W.; {Cuzzi}, J.N.
\newblock {Closed-form expressions for particle relative velocities induced by
  turbulence}.
\newblock {\em Astron. Astrophys.} {\bf 2007}, {\em 466},~413--420,
\newblock
  doi:{\changeurlcolor{black}\href{https://doi.org/10.1051/0004-6361:20066899}{\detokenize{10.1051/0004-6361:20066899}}}.

\bibitem[{Weidenschilling}(1977)]{Weidenschilling1977}
{Weidenschilling}, S.J.
\newblock {The distribution of mass in the planetary system and solar nebula}.
\newblock {\em Astrophys. Space Sci.} {\bf 1977}, {\em 51},~153--158,
\newblock
  doi:{\changeurlcolor{black}\href{https://doi.org/10.1007/BF00642464}{\detokenize{10.1007/BF00642464}}}.

\bibitem[{Deckers} and {Teiser}(2013)]{Deckers2013}
\textls[-25]{{Deckers}, J.; {Teiser}, J.
\newblock {Colliding Decimeter Dust}.
\newblock {\em Astrophys. J.} {\bf 2013}, {\em 769},~151,
\newblock
  doi:{\changeurlcolor{black}\href{https://doi.org/10.1088/0004-637X/769/2/151}{\detokenize{10.1088/0004-637X/769/2/151}}}.}

\bibitem[{Beitz} {et~al.}(2011){Beitz}, {G{\"u}ttler}, {Blum}, {Meisner},
  {Teiser}, and {Wurm}]{Beitz2011}
{Beitz}, E.; {G{\"u}ttler}, C.; {Blum}, J.; {Meisner}, T.; {Teiser}, J.;
  {Wurm}, G.
\newblock {Low-velocity Collisions of Centimeter-sized Dust Aggregates}.
\newblock {\em Astrophys. J.} {\bf 2011}, {\em 736},~34,
\newblock
  doi:{\changeurlcolor{black}\href{https://doi.org/10.1088/0004-637X/736/1/34}{\detokenize{10.1088/0004-637X/736/1/34}}}.

\bibitem[{Schr{\"a}pler} {et~al.}(2012){Schr{\"a}pler}, {Blum}, {Seizinger},
  and {Kley}]{Schraepler2012}
{Schr{\"a}pler}, R.; {Blum}, J.; {Seizinger}, A.; {Kley}, W.
\newblock {The Physics of Protoplanetesimal Dust Agglomerates. VII. The Low-
  velocity Collision Behavior of Large Dust Agglomerates}.
\newblock {\em Astrophys. J.} {\bf 2012}, {\em 758},~35,
\newblock
  doi:{\changeurlcolor{black}\href{https://doi.org/10.1088/0004-637X/758/1/35}{\detokenize{10.1088/0004-637X/758/1/35}}}.

\bibitem[{Birnstiel} {et~al.}(2010){Birnstiel}, {Dullemond}, and
  {Brauer}]{Birnstiel2010}
\textls[-5]{{Birnstiel}, T.; {Dullemond}, C.P.; {Brauer}, F.
\newblock {Gas- and dust evolution in protoplanetary disks}.
\newblock {\em Astron.~Astrophys.} {\bf 2010}, {\em 513},~A79,
\newblock
  doi:{\changeurlcolor{black}\href{https://doi.org/10.1051/0004-6361/200913731}{\detokenize{10.1051/0004-6361/200913731}}}.}

\bibitem[{Birnstiel} {et~al.}(2012){Birnstiel}, {Klahr}, and
  {Ercolano}]{Birnstiel2012}
\textls[-25]{{Birnstiel}, T.; {Klahr}, H.; {Ercolano}, B.
\newblock {A simple model for the evolution of the dust population in
  protoplanetary~disks}.
\newblock {\em Astron. Astrophys.} {\bf 2012}, {\em 539},~A148,
\newblock
  doi:{\changeurlcolor{black}\href{https://doi.org/10.1051/0004-6361/201118136}{\detokenize{10.1051/0004-6361/201118136}}}.}

\bibitem[{Wurm} {et~al.}(2005){Wurm}, {Paraskov}, and {Krauss}]{Wurm2005}
{Wurm}, G.; {Paraskov}, G.; {Krauss}, O.
\newblock {Growth of planetesimals by impacts at 25 m/s}.
\newblock {\em Icarus} {\bf 2005}, {\em 178},~253--263,
\newblock
  doi:{\changeurlcolor{black}\href{https://doi.org/10.1016/j.Icarus.2005.04.002}{\detokenize{10.1016/j.Icarus.2005.04.002}}}.

\bibitem[{Teiser} and {Wurm}(2009)]{Teiser2009a}
{Teiser}, J.; {Wurm}, G.
\newblock {High-velocity dust collisions: Forming planetesimals in a
  fragmentation cascade with final accretion}.
\newblock {\em Mon. Not. R. Astron. Soc.} {\bf 2009},
  {\em 393},~1584--1594,
\newblock
  doi:{\changeurlcolor{black}\href{https://doi.org/10.1111/j.1365-2966.2008.14289.x}{\detokenize{10.1111/j.1365-2966.2008.14289.x}}}.

\bibitem[{Meisner} {et~al.}(2013){Meisner}, {Wurm}, {Teiser}, and
  {Schywek}]{Meisner2013}
{Meisner}, T.; {Wurm}, G.; {Teiser}, J.; {Schywek}, M.
\newblock {Preplanetary scavengers: Growing tall in dust collisions}.
\newblock {\em Astron. Astrophys.} {\bf 2013}, {\em 559},~A123,
\newblock
  doi:{\changeurlcolor{black}\href{https://doi.org/10.1051/0004-6361/201322083}{\detokenize{10.1051/0004-6361/201322083}}}.

\bibitem[{Windmark} {et~al.}(2012){Windmark}, {Birnstiel}, {G{\"u}ttler},
  {Blum}, {Dullemond}, and {Henning}]{Windmark2012}
{Windmark}, F.; {Birnstiel}, T.; {G{\"u}ttler}, C.; {Blum}, J.; {Dullemond},
  C.P.; {Henning}, T.
\newblock {Planetesimal formation by sweep-up: How the bouncing barrier can be
  beneficial to growth}.
\newblock {\em Astron. Astrophys.} {\bf 2012}, {\em 540},~A73,
\newblock
  doi:{\changeurlcolor{black}\href{https://doi.org/10.1051/0004-6361/201118475}{\detokenize{10.1051/0004-6361/201118475}}}.

\bibitem[{Dr{\k{a}}{\.z}kowska} {et~al.}(2014){Dr{\k{a}}{\.z}kowska},
  {Windmark}, and {Dullemond}]{Drazkowska2014}
{Dr{\k{a}}{\.z}kowska}, J.; {Windmark}, F.; {Dullemond}, C.P.
\newblock {Modeling dust growth in protoplanetary disks: The breakthrough
  case}.
\newblock {\em Astron. Astrophys.} {\bf 2014}, {\em 567},~A38,
\newblock
  doi:{\changeurlcolor{black}\href{https://doi.org/10.1051/0004-6361/201423708}{\detokenize{10.1051/0004-6361/201423708}}}.

\bibitem[{Youdin} and {Goodman}(2005)]{Youdin2005}
{Youdin}, A.N.; {Goodman}, J.
\newblock {Streaming Instabilities in Protoplanetary Disks}.
\newblock {\em Astrophys. J.} {\bf 2005}, {\em 620},~459--469,
\newblock
  doi:{\changeurlcolor{black}\href{https://doi.org/10.1086/426895}{\detokenize{10.1086/426895}}}.

\bibitem[{Simon} {et~al.}(2016){Simon}, {Armitage}, {Li}, and
  {Youdin}]{Simon2016}
\textls[-25]{{Simon}, J.B.; {Armitage}, P.J.; {Li}, R.; {Youdin}, A.N.
\newblock {The Mass and Size Distribution of Planetesimals Formed by the
  Streaming Instability. I. The Role of Self-gravity}.
\newblock {\em Astrophys. J.} {\bf 2016}, {\em 822},~55,
\newblock
  doi:{\changeurlcolor{black}\href{https://doi.org/10.3847/0004-637X/822/1/55}{\detokenize{10.3847/0004-637X/822/1/55}}}.}

\bibitem[{Schneider} {et~al.}(2018){Schneider}, {Wurm}, {Teiser}, {Klahr},
  and {Carpenter}]{Schneider2018}
{Schneider}, N.; {Wurm}, G.; {Teiser}, J.; {Klahr}, H.; {Carpenter}, V.
\newblock {Streaming in laboratory experiments}.
\newblock {\em Astrophys.~J.} {\bf 2018}, submitted.

\bibitem[{Spohn}(2001)]{Spohn2001}
{Spohn}, T. {Planetologie}.
\newblock In {\em Erde und Planeten}; {Raith, W.}, Ed.; Walter de Gruyter: Berlin, Germany,
  2001; p.~427ff.

\bibitem[{Trieloff} and {Palme}(2006)]{Trieloff2006}
{Trieloff}, M.; {Palme}, H. {The Origin of Solids in the Early Solar System}.
\newblock In {\em Planet Formation}; {Klahr, H., Brandner,~W.}, Ed.;
  Cambridge University Press: Cambridge, UK, 2006; p.~64.

\bibitem[{Rappaport} {et~al.}(2013){Rappaport}, {Sanchis-Ojeda}, {Rogers},
  {Levine}, and {Winn}]{Rappaport2013}
{Rappaport}, S.; {Sanchis-Ojeda}, R.; {Rogers}, L.A.; {Levine}, A.; {Winn},
  J.N.
\newblock {The Roche Limit for Close-orbiting Planets: Minimum Density,
  Composition Constraints, and Application to the 4.2 hr Planet KOI 1843.03}.
\newblock {\em Astrophys. J.} {\bf 2013}, {\em 773},~L15,
\newblock
  doi:{\changeurlcolor{black}\href{https://doi.org/10.1088/2041-8205/773/1/L15}{\detokenize{10.1088/2041-8205/773/1/L15}}}.

\bibitem[{Sinukoff} {et~al.}(2017){Sinukoff}, {Howard}, {Petigura},
  {Fulton}, {Crossfield}, {Isaacson}, {Gonzales}, {Crepp}, {Brewer}, {Hirsch},
  {Weiss}, {Ciardi}, {Schlieder}, {Benneke}, {Christiansen}, {Dressing},
  {Hansen}, {Knutson}, {Kosiarek}, {Livingston}, {Greene}, {Rogers}, and
  {L{\'e}pine}]{Sinukoff2017}
{Sinukoff}, E.; {Howard}, A.W.; {Petigura}, E.A.; {Fulton}, B.J.; {Crossfield},
  I.J.M.; {Isaacson}, H.; {Gonzales}, E.; {Crepp},~J.R.; {Brewer}, J.M.;
  {Hirsch}, L.; et al.
\newblock {K2-66b and K2-106b: Two Extremely Hot Sub-Neptune-size Planets with
  High Densities}.
\newblock {\em Astron. J.} {\bf 2017}, {\em 153},~271,
\newblock
  doi:{\changeurlcolor{black}\href{https://doi.org/10.3847/1538-3881/aa725f}{\detokenize{10.3847/1538-3881/aa725f}}}.

\bibitem[{Santerne} {et~al.}(2018){Santerne}, {Brugger}, {Armstrong},
  {Adibekyan}, {Lillo-Box}, {Gosselin}, {Aguichine}, {Almenara}, {Barrado},
  {Barros}, {Bayliss}, {Boisse}, {Bonomo}, {Bouchy}, {Brown}, {Deleuil},
  {Delgado Mena}, {Demangeon}, {D{\'{\i}}az}, {Doyle}, {Dumusque}, {Faedi},
  {Faria}, {Figueira}, {Foxell}, {Giles}, {H{\'e}brard}, {Hojjatpanah},
  {Hobson}, {Jackman}, {King}, {Kirk}, {Lam}, {Ligi}, {Lovis}, {Louden},
  {McCormac}, {Mousis}, {Neal}, {Osborn}, {Pepe}, {Pollacco}, {Santos},
  {Sousa}, {Udry}, and {Vigan}]{Santerne2018}
{Santerne}, A.; {Brugger}, B.; {Armstrong}, D.J.; {Adibekyan}, V.; {Lillo-Box},
  J.; {Gosselin}, H.; {Aguichine}, A.; {Almenara}, J.M.; {Barrado}, D.;
  {Barros}, S.C.C.; et al.
\newblock {An Earth-sized exoplanet with a Mercury-like composition}.
\newblock {\em Nat.~Astron.} {\bf 2018}, {\em 2},~393--400,
\newblock
  doi:{\changeurlcolor{black}\href{https://doi.org/10.1038/s41550-018-0420-5}{\detokenize{10.1038/s41550-018-0420-5}}}.

\bibitem[{Guenther} {et~al.}(2017){Guenther}, {Barrag{\'a}n}, {Dai},
  {Gandolfi}, {Hirano}, {Fridlund}, {Fossati}, {Chau}, {Helled}, {Korth},
  {Prieto-Arranz}, {Nespral}, {Antoniciello}, {Deeg}, {Hjorth}, {Grziwa},
  {Albrecht}, {Hatzes}, {Rauer}, {Csizmadia}, {Smith}, {Cabrera}, {Narita},
  {Arriagada}, {Burt}, {Butler}, {Cochran}, {Crane}, {Eigm{\"u}ller},
  {Erikson}, {Johnson}, {Kiilerich}, {Kubyshkina}, {Palle}, {Persson},
  {P{\"a}tzold}, {Sabotta}, {Sato}, {Shectman}, {Teske}, {Thompson}, {Van
  Eylen}, {Nowak}, {Vanderburg}, {Winn}, and {Wittenmyer}]{Guenther2017}
{Guenther}, E.W.; {Barrag{\'a}n}, O.; {Dai}, F.; {Gandolfi}, D.; {Hirano}, T.;
  {Fridlund}, M.; {Fossati}, L.; {Chau}, A.; {Helled}, R.; {Korth}, J.;
 et al.
\newblock {K2-106, a system containing a metal-rich planet and a planet of
  lower density}.
\newblock {\em Astron.~Astrophys.} {\bf 2017}, {\em 608},~A93,
\newblock
  doi:{\changeurlcolor{black}\href{https://doi.org/10.1051/0004-6361/201730885}{\detokenize{10.1051/0004-6361/201730885}}}.

\bibitem[{Benz} {et~al.}(1988){Benz}, {Slattery}, and {Cameron}]{Benz1988}
{Benz}, W.; {Slattery}, W.L.; {Cameron}, A.G.W.
\newblock {Collisional stripping of Mercury's mantle}.
\newblock {\em Icarus} {\bf 1988}, {\em 74},~516--528,
\newblock
  doi:{\changeurlcolor{black}\href{https://doi.org/10.1016/0019-1035(88)90118-2}{\detokenize{10.1016/0019-1035(88)90118-2}}}.

\bibitem[{Cameron}(1985)]{Cameron1985}
\textls[-25]{{Cameron}, A.G.W.
\newblock {The partial volatilization of Mercury}.
\newblock {\em Icarus} {\bf 1985}, {\em 64},~285--294,
\newblock
  doi:{\changeurlcolor{black}\href{https://doi.org/10.1016/0019-1035(85)90091-0}{\detokenize{10.1016/0019-1035(85)90091-0}}}.}

\bibitem[{Peplowski} {et~al.}(2011){Peplowski}, {Blewett}, {Denevi},
  {Evans}, {Lawrence}, {Nittler}, {Rhodes}, {Selby}, and
  {Solomon}]{Peplowski2011}
{Peplowski}, P.N.; {Blewett}, D.T.; {Denevi}, B.W.; {Evans}, L.G.; {Lawrence},
  D.J.; {Nittler}, L.R.; {Rhodes}, E.A.; {Selby},~C.M.; {Solomon}, S.C.
\newblock {Mapping iron abundances on the surface of Mercury: Predicted spatial
  resolution of the MESSENGER Gamma-Ray Spectrometer}.
\newblock {\em Planet. Space Sci.} {\bf 2011}, {\em 59},~1654--1658,
\newblock
  doi:{\changeurlcolor{black}\href{https://doi.org/10.1016/j.pss.2011.06.001}{\detokenize{10.1016/j.pss.2011.06.001}}}.

\bibitem[{Gundlach} and {Blum}(2015)]{Gundlach2015}
{Gundlach}, B.; {Blum}, J.
\newblock {The Stickiness of Micrometer-sized Water-ice Particles}.
\newblock {\em Astrophys. J.} {\bf 2015}, {\em 798},~34,
\newblock
  doi:{\changeurlcolor{black}\href{https://doi.org/10.1088/0004-637X/798/1/34}{\detokenize{10.1088/0004-637X/798/1/34}}}.

\bibitem[{Deckers} and {Teiser}(2016)]{Deckers2016}
{Deckers}, J.; {Teiser}, J.
\newblock {Collisions of solid ice in planetesimal formation}.
\newblock {\em Mon. Not. R. Astron. Soc.} {\bf 2016},
  {\em 456},~4328--4334,
\newblock doi:{\changeurlcolor{black}\href{https://doi.org/10.1093/mnras/stv2952}{\detokenize{10.1093/mnras/stv2952}}}.

\bibitem[{Musiolik} {et~al.}(2016){Musiolik}, {Teiser}, {Jankowski}, and
  {Wurm}]{Musiolik2016}
{Musiolik}, G.; {Teiser}, J.; {Jankowski}, T.; {Wurm}, G.
\newblock {Ice Grain Collisions in Comparison: CO$_2$, H$_2$O, and Their~Mixtures}.
\newblock {\em Astrophys. J.} {\bf 2016}, {\em 827},~63,
\newblock
  doi:{\changeurlcolor{black}\href{https://doi.org/10.3847/0004-637X/827/1/63}{\detokenize{10.3847/0004-637X/827/1/63}}}.

\bibitem[{Kataoka} {et~al.}(2013){Kataoka}, {Tanaka}, {Okuzumi}, and
  {Wada}]{Kataoka2013}
{Kataoka}, A.; {Tanaka}, H.; {Okuzumi}, S.; {Wada}, K.
\newblock {Fluffy dust forms icy planetesimals by static compression}.
\newblock {\em Astron. Astrophys.} {\bf 2013}, {\em 557},~L4,
\newblock
  doi:{\changeurlcolor{black}\href{https://doi.org/10.1051/0004-6361/201322151}{\detokenize{10.1051/0004-6361/201322151}}}.

\bibitem[{Okuzumi} {et~al.}(2012){Okuzumi}, {Tanaka}, {Kobayashi}, and
  {Wada}]{Okuzumi2012x}
{Okuzumi}, S.; {Tanaka}, H.; {Kobayashi}, H.; {Wada}, K.
\newblock {Rapid Coagulation of Porous Dust Aggregates outside the Snow Line: A
  Pathway to Successful Icy Planetesimal Formation}.
\newblock {\em Astrophys. J.} {\bf 2012}, {\em 752},~106,
\newblock
  doi:{\changeurlcolor{black}\href{https://doi.org/10.1088/0004-637X/752/2/106}{\detokenize{10.1088/0004-637X/752/2/106}}}.

\bibitem[{Musiolik} {et~al.}(2016){Musiolik}, {Teiser}, {Jankowski}, and
  {Wurm}]{Musiolik2016b}
{Musiolik}, G.; {Teiser}, J.; {Jankowski}, T.; {Wurm}, G.
\newblock {Collisions of CO$_2$ Ice Grains in Planet Formation}.
\newblock {\em Astrophys.~J.} {\bf 2016}, {\em 818},~16,
\newblock
  doi:{\changeurlcolor{black}\href{https://doi.org/10.3847/0004-637X/818/1/16}{\detokenize{10.3847/0004-637X/818/1/16}}}.

\bibitem[{Pinilla} {et~al.}(2017){Pinilla}, {Pohl}, {Stammler}, and
  {Birnstiel}]{Pinilla2017}
{Pinilla}, P.; {Pohl}, A.; {Stammler}, S.M.; {Birnstiel}, T.
\newblock {Dust Density Distribution and Imaging Analysis of Different Ice
  Lines in Protoplanetary Disks}.
\newblock {\em Astrophys. J.} {\bf 2017}, {\em 845},~68,
\newblock
  doi:{\changeurlcolor{black}\href{https://doi.org/10.3847/1538-4357/aa7edb}{\detokenize{10.3847/1538-4357/aa7edb}}}.

\bibitem[{Sch{\"a}fer} {et~al.}(2016){Sch{\"a}fer}, {Riecker}, {Maindl},
  {Speith}, {Scherrer}, and {Kley}]{Schaefer2016}
{Sch{\"a}fer}, C.; {Riecker}, S.; {Maindl}, T.I.; {Speith}, R.; {Scherrer}, S.;
  {Kley}, W.
\newblock {A smooth particle hydrodynamics code to model collisions between
  solid, self-gravitating objects}.
\newblock {\em Astron. Astrophys.} {\bf 2016}, {\em 590},~A19,
\newblock
  doi:{\changeurlcolor{black}\href{https://doi.org/10.1051/0004-6361/201528060}{\detokenize{10.1051/0004-6361/201528060}}}.

\bibitem[{Geretshauser} {et~al.}(2011{\natexlab{a}}){Geretshauser}, {Meru},
  {Speith}, and {Kley}]{Geretshauser2011}
{Geretshauser}, R.J.; {Meru}, F.; {Speith}, R.; {Kley}, W.
\newblock {The four-population model: A new classification scheme for
  pre-planetesimal collisions}.
\newblock {\em Astron. Astrophys.} {\bf 2011}, {\em 531},~A166,
\newblock
  doi:{\changeurlcolor{black}\href{https://doi.org/10.1051/0004-6361/201116901}{\detokenize{10.1051/0004-6361/201116901}}}.

\bibitem[{Geretshauser} {et~al.}(2011{\natexlab{b}}){Geretshauser},
  {Speith}, and {Kley}]{Geretshauser2011a}
{Geretshauser}, R.J.; {Speith}, R.; {Kley}, W.
\newblock {Collisions of inhomogeneous pre-planetesimals}.
\newblock {\em Astron. Astrophys.} {\bf 2011}, {\em 536},~A104,
\newblock
  doi:{\changeurlcolor{black}\href{https://doi.org/10.1051/0004-6361/201117645}{\detokenize{10.1051/0004-6361/201117645}}}.

\bibitem[{Blum} and {Schraepler}(2004)]{Blum2004}
\textls[-5]{{Blum}, J.; {Schraepler}, R.
\newblock {Structure and Mechanical Properties of High-Porosity Macroscopic
  Agglomerates Formed by Random Ballistic Deposition}.
\newblock {\em Phys. Rev. Lett.} {\bf 2004}, {\em 93},~115503,
\newblock
  doi:{\changeurlcolor{black}\href{https://doi.org/10.1103/PhysRevLett.93.115503}{\detokenize{10.1103/PhysRevLett.93.115503}}}.}

\bibitem[{Meisner} {et~al.}(2012){Meisner}, {Wurm}, and
  {Teiser}]{Meisner2012}
{Meisner}, T.; {Wurm}, G.; {Teiser}, J.
\newblock {Experiments on centimeter-sized dust aggregates and their
  implications for planetesimal formation}.
\newblock {\em Astron. Astrophys.} {\bf 2012}, {\em 544},~A138,
\newblock
  doi:{\changeurlcolor{black}\href{https://doi.org/10.1051/0004-6361/201219099}{\detokenize{10.1051/0004-6361/201219099}}}.

\bibitem[{de Beule} {et~al.}(2017){de Beule}, {Landers}, {Salamon}, {Wende},
  and {Wurm}]{DeBeule2017}
{De Beule}, C.; {Landers}, J.; {Salamon}, S.; {Wende}, H.; {Wurm}, G.
\newblock {Planetesimal Formation in the Warm, Inner Disk: Experiments with
  Tempered Dust}.
\newblock {\em Astrophys. J.} {\bf 2017}, {\em 837},~59,
\newblock
  doi:{\changeurlcolor{black}\href{https://doi.org/10.3847/1538-4357/837/1/59}{\detokenize{10.3847/1538-4357/837/1/59}}}.

\bibitem[{Zanda}(2004)]{Zanda2004}
{Zanda}, B.
\newblock {Chondrules}.
\newblock {\em Earth Planet. Sci. Lett.} {\bf 2004}, {\em
  224},~1--17,
\newblock
  doi:{\changeurlcolor{black}\href{https://doi.org/10.1016/j.epsl.2004.05.005}{\detokenize{10.1016/j.epsl.2004.05.005}}}.

\bibitem[{Connolly} {et~al.}(1994){Connolly}, {Hewins}, {Atre}, and
  {Lofgren}]{Connolly1994}
\textls[-15]{{Connolly}, Jr., H.C.; {Hewins}, R.H.; {Atre}, N.; {Lofgren}, G.E.
\newblock {Compound chondrules: An experimental~investigation}.
\newblock {\em Meteoritics} {\bf 1994}, {\em 29},~458.}

\bibitem[{Bodgan} {et~al.}(2018){Bodgan}, {Teiser}, {Fischer}, {Kruss}, and
  {Wurm}]{Bogdan2018}
{Bodgan}, T.; {Teiser}, J.; {Fischer}, N.; {Kruss}, M.; {Wurm}, G.
\newblock {Constrains on compound chondrule formation from laboratory high
  temperature collisions}.
\newblock {\em Icarus} {\bf 2018}, submitted.

\bibitem[{Krauss} and {Wurm}(2005)]{Krauss2005}
{Krauss}, O.; {Wurm}, G.
\newblock Photophoresis and the Pile-up of Dust in Young Circumstellar Disks.
\newblock {\em Astrophys J.} {\bf 2005}, {\em 630},~1088--1092,
\newblock
  doi:{\changeurlcolor{black}\href{https://doi.org/10.1086/432087}{\detokenize{10.1086/432087}}}.

\bibitem[{Wurm} and {Krauss}(2006)]{Wurm2006}
{Wurm}, G.; {Krauss}, O.
\newblock {Concentration and sorting of chondrules and CAIs in the late Solar
  Nebula}.
\newblock {\em Icarus} {\bf 2006}, {\em 180},~487--495,
\newblock
  doi:{\changeurlcolor{black}\href{https://doi.org/10.1016/j.Icarus.2005.09.011}{\detokenize{10.1016/j.Icarus.2005.09.011}}}.

\bibitem[{Herrmann} and {Krivov}(2007)]{Herrmann2007}
{Herrmann}, F.; {Krivov}, A.V.
\newblock {Effects of photophoresis on the evolution of transitional
  circumstellar disks}.
\newblock {\em Astron. Astrophys.} {\bf 2007}, {\em 476},~829--839,
  \href{http://xxx.lanl.gov/abs/0711.2595}.
\newblock
  doi:{\changeurlcolor{black}\href{https://doi.org/10.1051/0004-6361:20078322}{\detokenize{10.1051/0004-6361:20078322}}}.

\bibitem[{Takeuchi} and {Krauss}(2008)]{Takeuchi2008}
\textls[-25]{{Takeuchi}, T.; {Krauss}, O.
\newblock {Photophoretic Structuring of Circumstellar Dust Disks}.
\newblock {\em Astrophys. J.} {\bf 2008}, {\em 677},~1309--1323,
  doi:{\changeurlcolor{black}\href{https://doi.org/10.1086/527426}{\detokenize{10.1086/527426}}}.
}
\bibitem[{Rohatschek}(1995)]{Rohatschek1995}
{Rohatschek}, H.
\newblock Semi-empirical model of photophoretic forces for the entire range of
  pressures.
\newblock {\em J. Aerosol Sci.} {\bf 1995}, {\em 26},~717--734,
\newblock
  doi:{\changeurlcolor{black}\href{https://doi.org/10.1016/0021-8502(95)00011-Z}{\detokenize{10.1016/0021-8502(95)00011-Z}}}.

\bibitem[{Cordier} {et~al.}(2016){Cordier}, {Prada Moroni}, and
  {Tognelli}]{Cordier2016}
{Cordier}, D.; {Prada Moroni}, P.G.; {Tognelli}, E.
\newblock {Dust photophoretic transport around a T Tauri star: Implications for
  comets composition}.
\newblock {\em Icarus} {\bf 2016}, {\em 268},~281--294,
\newblock
  doi:{\changeurlcolor{black}\href{https://doi.org/10.1016/j.Icarus.2015.11.037}{\detokenize{10.1016/j.Icarus.2015.11.037}}}.

\bibitem[{Beresnev} {et~al.}(2003){Beresnev}, {Kovalev}, {Kochneva},
  {Runkov}, {Suetin}, and {Cheremisin}]{Beresnev2003zu}
{Beresnev}, S.A.; {Kovalev}, F.D.; {Kochneva}, L.B.; {Runkov}, V.A.; {Suetin},
  P.E.; {Cheremisin}, A.A.
\newblock {On the possibility of particle's photophoretic levitation in the
  stratosphere}.
\newblock {\em Atmos. Ocean. Opt.}
  {\bf 2003}, {\em 16},~44--48.
%\newblock Translated by {Terpugova}, S.~A. and edited by {Arshinov}, Y.~F.

\bibitem[{Matthews} {et~al.}(2016){Matthews}, {Kimery}, {Wurm}, {de Beule},
  {Kuepper}, and {Hyde}]{Matthews2016}
{Matthews}, L.S.; {Kimery}, J.B.; {Wurm}, G.; {de Beule}, C.; {Kuepper}, M.;
  {Hyde}, T.W.
\newblock {Photophoretic force on aggregate grains}.
\newblock {\em Mon. Not. R. Astron. Soc.} {\bf 2016},
  {\em 455},~2582--2591,
\newblock doi:{\changeurlcolor{black}\href{https://doi.org/10.1093/mnras/stv2532}{\detokenize{10.1093/mnras/stv2532}}}.

\bibitem[{Loesche} and {Wurm}(2012)]{Loesche2012}
{Loesche}, C.; {Wurm}, G.
\newblock {Thermal and photophoretic properties of dust mantled chondrules and
  sorting in the solar nebula}.
\newblock {\em Astron. Astrophys.} {\bf 2012}, {\em 545},~A36,
\newblock
  doi:{\changeurlcolor{black}\href{https://doi.org/10.1051/0004-6361/201218989}{\detokenize{10.1051/0004-6361/201218989}}}.

\bibitem[{Wurm} {et~al.}(2013){Wurm}, {Trieloff}, and {Rauer}]{Wurm2013}
\textls[-25]{{Wurm}, G.; {Trieloff}, M.; {Rauer}, H.
\newblock {Photophoretic Separation of Metals and Silicates: The Formation of
  Mercury-like Planets and Metal Depletion in Chondrites}.
\newblock {\em Astrophys. J.} {\bf 2013}, {\em 769},~78,
\newblock
  doi:{\changeurlcolor{black}\href{https://doi.org/10.1088/0004-637X/769/1/78}{\detokenize{10.1088/0004-637X/769/1/78}}}.}

\bibitem[{Cuello} {et~al.}(2016){Cuello}, {Gonzalez}, and
  {Pignatale}]{Cuello2016}
{Cuello}, N.; {Gonzalez}, J.F.; {Pignatale}, F.C.
\newblock {Effects of photophoresis on the dust distribution in a 3D
  protoplanetary disc}.
\newblock {\em Mon. Not. R. Astron. Soc.} {\bf 2016},
  {\em 458},~2140--2149,
\newblock doi:{\changeurlcolor{black}\href{https://doi.org/10.1093/mnras/stw396}{\detokenize{10.1093/mnras/stw396}}}.

\bibitem[{Krauss} {et~al.}(2007){Krauss}, {Wurm}, {Mousis}, {Petit},
  {Horner}, and {Alibert}]{Krauss2007}
{Krauss}, O.; {Wurm}, G.; {Mousis}, O.; {Petit}, J.M.; {Horner}, J.; {Alibert},
  Y.
\newblock {The photophoretic sweeping of dust in transient protoplanetary
  disks}.
\newblock {\em Astron. Astrophys.} {\bf 2007}, {\em 462},~977--987,
\newblock
  doi:{\changeurlcolor{black}\href{https://doi.org/10.1051/0004-6361:20066363}{\detokenize{10.1051/0004-6361:20066363}}}.

\bibitem[{Mousis} {et~al.}(2007){Mousis}, {Petit}, {Wurm}, {Krauss},
  {Alibert}, and {Horner}]{Mousis2007}
{Mousis}, O.; {Petit}, J.M.; {Wurm}, G.; {Krauss}, O.; {Alibert}, Y.; {Horner},
  J.
\newblock Photophoresis as a source of hot minerals in comets.
\newblock {\em Astron. Astrophys.} {\bf 2007}, {\em 466},~L9--L12,
\newblock
  doi:{\changeurlcolor{black}\href{https://doi.org/10.1051/0004-6361:20077170}{\detokenize{10.1051/0004-6361:20077170}}}.

\bibitem[{Moudens} {et~al.}(2011){Moudens}, {Mousis}, {Petit}, {Wurm},
  {Cordier}, and {Charnoz}]{Moudens2011}
{Moudens}, A.; {Mousis}, O.; {Petit}, J.M.; {Wurm}, G.; {Cordier}, D.;
  {Charnoz}, S.
\newblock {Photophoretic transport of hot minerals in the solar nebula}.
\newblock {\em Astron. Astrophys.} {\bf 2011}, {\em 531},~A106,
\newblock
  doi:{\changeurlcolor{black}\href{https://doi.org/10.1051/0004-6361/201116476}{\detokenize{10.1051/0004-6361/201116476}}}.

\bibitem[{Haack} and {Wurm}(2007)]{Haack2007}
{Haack}, H.; {Wurm}, G.
\newblock Life on the Edge---Formation of CAIs and Chondrules at the Inner Edge
  of the Dust~Disk. {\em Meteorit. Planet. Sci. Suppl.}
\newblock  2007,~42, 5157.

\bibitem[{Wurm} and {Haack}(2009)]{Wurm2009}
{Wurm}, G.; {Haack}, H.
\newblock {Outward transport of CAIs during FU-Orionis events}.
\newblock {\em Meteorit. Planet. Sci.} {\bf 2009}, {\em
  44},~689--699,
\newblock
  doi:{\changeurlcolor{black}\href{https://doi.org/10.1111/j.1945-5100.2009.tb00763.x}{\detokenize{10.1111/j.1945-5100.2009.tb00763.x}}}.

\bibitem[{McNally} and {McClure}(2017)]{McNally2017}
{McNally}, C.P.; {McClure}, M.K.
\newblock {Photophoretic Levitation and Trapping of Dust in the Inner Regions
  of Protoplanetary Disks}.
\newblock {\em Astrophys. J.} {\bf 2017}, {\em 834},~48,
\newblock
  doi:{\changeurlcolor{black}\href{https://doi.org/10.3847/1538-4357/834/1/48}{\detokenize{10.3847/1538-4357/834/1/48}}}.

\bibitem[{Loesche} {et~al.}(2016){Loesche}, {Wurm}, {Kelling}, {Teiser}, and
  {Ebel}]{Loesche2016}
{Loesche}, C.; {Wurm}, G.; {Kelling}, T.; {Teiser}, J.; {Ebel}, D.S.
\newblock {The motion of chondrules and other particles in a protoplanetary
  disc with temperature fluctuations}.
\newblock {\em Mon. Not. R. Astron. Soc.} {\bf 2016},
  {\em 463},~4167--4174,
\newblock doi:{\changeurlcolor{black}\href{https://doi.org/10.1093/mnras/stw2279}{\detokenize{10.1093/mnras/stw2279}}}.

\bibitem[{McNally} and {Hubbard}(2015)]{McNally2015}
{McNally}, C.P.; {Hubbard}, A.
\newblock {Photophoresis in a Dilute, Optically Thick Medium and Dust Motion in
  Protoplanetary Disks}.
\newblock {\em Astrophys. J.} {\bf 2015}, {\em 814},~37,
\newblock
  doi:{\changeurlcolor{black}\href{https://doi.org/10.1088/0004-637X/814/1/37}{\detokenize{10.1088/0004-637X/814/1/37}}}.

\bibitem[{Wurm} {et~al.}(2010){Wurm}, {Teiser}, {Bischoff}, {Haack}, and
  {Roszjar}]{Wurm2010}
{Wurm}, G.; {Teiser}, J.; {Bischoff}, A.; {Haack}, H.; {Roszjar}, J.
\newblock {Experiments on the photophoretic motion of chondrules and dust
  aggregates---Indications for the transport of matter in protoplanetary
  disks}.
\newblock {\em ICARUS} {\bf 2010}, {\em 208},~482--491,
\newblock
  doi:{\changeurlcolor{black}\href{https://doi.org/10.1016/j.Icarus.2010.01.033}{\detokenize{10.1016/j.Icarus.2010.01.033}}}.

\bibitem[{Loesche} {et~al.}(2014){Loesche}, {Teiser}, {Wurm}, {Hesse},
  {Friedrich}, and {Bischoff}]{Loesche2014}
{Loesche}, C.; {Teiser}, J.; {Wurm}, G.; {Hesse}, A.; {Friedrich}, J.M.;
  {Bischoff}, A.
\newblock {Photophoretic Strength on Chondrules. 2. Experiment}.
\newblock {\em Astrophys. J.} {\bf 2014}, {\em 792},~73,
\newblock
  doi:{\changeurlcolor{black}\href{https://doi.org/10.1088/0004-637X/792/1/73}{\detokenize{10.1088/0004-637X/792/1/73}}}.

\bibitem[{Kuepper} {et~al.}(2014){Kuepper}, de~{Beule}, {Wurm}, {Matthews},
  {Kimery}, and {Hyde}]{Kuepper2014}
{Kuepper}, M.; de~{Beule}, C.; {Wurm}, G.; {Matthews}, L.S.; {Kimery}, J.S.;
  {Hyde}, T.W.
\newblock {Photophoresis on polydisperse basalt microparticles under
  microgravity}.
\newblock {\em J. Aerosol Sci.} {\bf 2014}, {\em 76},~126--137,
\newblock
  doi:{\changeurlcolor{black}\href{https://doi.org/10.1016/j.jaerosci.2014.06.008}{\detokenize{10.1016/j.jaerosci.2014.06.008}}}.

\bibitem[{von Borstel} and {Blum}(2012)]{vanBorstel2012}
{von Borstel}, I.; {Blum}, J.
\newblock {Photophoresis of dust aggregates in protoplanetary disks}.
\newblock {\em Astron. Astrophys.} {\bf 2012}, {\em 548},~A96,
\newblock
  doi:{\changeurlcolor{black}\href{https://doi.org/10.1051/0004-6361/201219622}{\detokenize{10.1051/0004-6361/201219622}}}.

\bibitem[{Beresnev} {et~al.}(2003){Beresnev}, {Kochneva}, and
  {Suetin}]{Beresnev2003b}
{Beresnev}, S.A.; {Kochneva}, L.B.; {Suetin}, P.E.
\newblock {Photophoresis Of Aerosols In The Earth Atmosphere}.
\newblock {\em Thermophys.~Aeromech.} {\bf 2003}, {\em 10},~287--301.

\bibitem[{van Eymeren} and {Wurm}(2012)]{vanEymeren2012}
{van Eymeren}, J.; {Wurm}, G.
\newblock {The implications of particle rotation on the effect of
  photophoresis}.
\newblock {\em Mon. Not. R. Astron. Soc.} {\bf 2012},
  {\em 420},~183--186,
\newblock
  doi:{\changeurlcolor{black}\href{https://doi.org/10.1111/j.1365-2966.2011.20020.x}{\detokenize{10.1111/j.1365-2966.2011.20020.x}}}.

\bibitem[{Loesche} {et~al.}(2016){Loesche}, {Wurm}, {Jankowski}, and
  {Kuepper}]{Loesche2016a}
{Loesche}, C.; {Wurm}, G.; {Jankowski}, T.; {Kuepper}, M.
\newblock {Photophoresis on particles hotter/colder than the ambient gas in the
  free molecular flow}.
\newblock {\em {J. Aerosol Sci.}} {\bf 2016}, {\em 97},~22--33,
\newblock
  doi:{\changeurlcolor{black}\href{https://doi.org/10.1016/j.jaerosci.2016.04.001}{\detokenize{10.1016/j.jaerosci.2016.04.001}}}.

\bibitem[{Loesche} and {Husmann}(2016)]{Loesche2016c}
{Loesche}, C.; {Husmann}, T.
\newblock {Photophoresis on particles hotter/colder than the ambient gas for
  the entire range of pressures}.
\newblock {\em J. Aerosol Sci.} {\bf 2016}, {\em 102},~55--71,
\newblock
  doi:{\changeurlcolor{black}\href{https://doi.org/10.1016/j.jaerosci.2016.08.013}{\detokenize{10.1016/j.jaerosci.2016.08.013}}}.

\bibitem[{Cheremisin} {et~al.}(2005){Cheremisin}, {Vassilyev}, and
  {Horvath}]{Cheremisin2005}
{Cheremisin}, A.A.; {Vassilyev}, Y.V.; {Horvath}, H.
\newblock Gravito-photophoresis and aerosol stratification in the~atmosphere.
\newblock {\em J. Aerosol Sci.} {\bf 2005}, {\em 36},~1277--1299,
\newblock
  doi:{\changeurlcolor{black}\href{https://doi.org/10.1016/j.jaerosci.2005.02.003}{\detokenize{10.1016/j.jaerosci.2005.02.003}}}.

\bibitem[{Cheremisin} {et~al.}(2011){Cheremisin}, {Shnipov}, {Horvath}, and
  {Rohatschek}]{Cheremisin2011}
{Cheremisin}, A.A.; {Shnipov}, I.S.; {Horvath}, H.; {Rohatschek}, H.
\newblock {The global picture of aerosol layers formation in the stratosphere
  and in the mesosphere under the influence of gravito-photophoretic and
  magneto-photophoretic forces}.
\newblock {\em J. Geophys. Res. (Atmos.)} {\bf 2011}, {\em
  116},~19204,
\newblock
  doi:{\changeurlcolor{black}\href{https://doi.org/10.1029/2011JD015958}{\detokenize{10.1029/2011JD015958}}}.

\bibitem[Rohatschek(1956)]{Rohatschek1956b}
Rohatschek, H.
\newblock {{\"U}ber die Kr{\"a}fte der reinen Photophorese und der
  Gravitophotophorese (On the forces of pure and gravito-photophoresis)}.
\newblock {\em Acta Phys. Austriaca} {\bf 1956}, {\em 10},~267--286.

\bibitem[{Rohatschek}(1985)]{Rohatschek1985Exp}
{Rohatschek}, H.
\newblock Direction, magnitude and causes of photophoretic forces.
\newblock {\em J. Aerosol Sci.} {\bf 1985}, {\em 16},~29--42,
\newblock
  doi:{\changeurlcolor{black}\href{https://doi.org/10.1016/0021-8502(85)90018-7}{\detokenize{10.1016/0021-8502(85)90018-7}}}.

\bibitem[{Rohatschek}(1989)]{Rohatschek1989}
{Rohatschek}, H.
\newblock {Photophoretic levitation of carbonaceous aerosols}.
\newblock {\em J. Aerosol Sci.} {\bf 1989}, {\em 20},~903--906,
%\newblock Proceedings of the 1989 European Aerosol Research,
  doi:{\changeurlcolor{black}\href{https://doi.org/10.1016/0021-8502(89)90722-2}{\detokenize{10.1016/0021-8502(89)90722-2}}}.

\bibitem[{Wurm}(2007)]{Wurm2007x}
{Wurm}, G.
\newblock {Light-induced disassembly of dusty bodies in inner protoplanetary
  discs: Implications for the formation of planets}.
\newblock {\em Mon. Not. R. Astron. Soc.} {\bf 2007},
  {\em 380},~683--690,  
\newblock
  doi:{\changeurlcolor{black}\href{https://doi.org/10.1111/j.1365-2966.2007.12105.x}{\detokenize{10.1111/j.1365-2966.2007.12105.x}}}.

\bibitem[{Wurm} and {Krauss}(2006)]{Wurm2006a}
{Wurm}, G.; {Krauss}, O.
\newblock {Dust Eruptions by Photophoresis and Solid State Greenhouse Effects}.
\newblock {\em Phys. Rev. Lett.} {\bf 2006}, {\em 96},~134301,
\newblock
  doi:{\changeurlcolor{black}\href{https://doi.org/10.1103/PhysRevLett.96.134301}{\detokenize{10.1103/PhysRevLett.96.134301}}}.

\bibitem[{de Beule} {et~al.}(2013){de Beule}, {Kelling}, {Wurm}, {Teiser},
  and {Jankowski}]{Beule2013}
{{De Beule}}, C.; {Kelling}, T.; {Wurm}, G.; {Teiser}, J.; {Jankowski}, T.
\newblock {From Planetesimals to Dust: Low-gravity Experiments on Recycling
  Solids at the Inner Edges of Protoplanetary Disks}.
\newblock {\em Astrophys. J.} {\bf 2013}, {\em 763},~11,
\newblock
  doi:{\changeurlcolor{black}\href{https://doi.org/10.1088/0004-637X/763/1/11}{\detokenize{10.1088/0004-637X/763/1/11}}}.%ref 127 and 130 are the same. .. removed130

\bibitem[{Kelling} and {Wurm}(2011)]{Kelling2011}
{Kelling}, T.; {Wurm}, G.
\newblock {A Mechanism to Produce the Small Dust Observed in Protoplanetary
  Disks}.
\newblock {\em Astrophys.~J.} {\bf 2011}, {\em 733},~120--125,
\newblock
  doi:{\changeurlcolor{black}\href{https://doi.org/10.1088/0004-637X/733/2/120}{\detokenize{10.1088/0004-637X/733/2/120}}}.

\bibitem[{Kocifaj} {et~al.}(2010){Kocifaj}, {Kla{\v c}ka}, {Wurm},
  {Kelling}, and {Koh{\'u}t}]{Kocifaj2010}
{Kocifaj}, M.; {Kla{\v c}ka}, J.; {Wurm}, G.; {Kelling}, T.; {Koh{\'u}t}, I.
\newblock {Dust ejection from (pre-)planetary bodies by temperature gradients:
  radiative and heat transfer}.
\newblock {\em Mon. Not. R. Astron. Soc.} {\bf 2010},
  {\em 404},~1512--1518,
\newblock
  doi:{\changeurlcolor{black}\href{https://doi.org/10.1111/j.1365-2966.2010.16370.x}{\detokenize{10.1111/j.1365-2966.2010.16370.x}}}.

\bibitem[{Schmidt} {et~al.}(2017){Schmidt}, {Andrieu}, {Costard}, {Kocifaj},
  and {Meresescu}]{Schmidt2017}
{Schmidt}, F.; {Andrieu}, F.; {Costard}, F.; {Kocifaj}, M.; {Meresescu}, A.G.
\newblock {Formation of recurring slope lineae on Mars by rarefied
  gas-triggered granular flows}.
\newblock {\em Nat. Geosci.} {\bf 2017}, {\em 10},~270--273,
\newblock
  doi:{\changeurlcolor{black}\href{https://doi.org/10.1038/ngeo2917}{\detokenize{10.1038/ngeo2917}}}.

\bibitem[{de Beule} {et~al.}(2015){de Beule}, {Wurm}, {Kelling}, {Koester},
  and {Kocifaj}]{Debeule2015x}
{de Beule}, C.; {Wurm}, G.; {Kelling}, T.; {Koester}, M.; {Kocifaj}, M.
\newblock {An insolation activated dust layer on Mars}.
\newblock {\em Icarus} {\bf 2015}, {\em 260},~23--28,
\newblock
  doi:{\changeurlcolor{black}\href{https://doi.org/10.1016/j.Icarus.2015.06.002}{\detokenize{10.1016/j.Icarus.2015.06.002}}}.

\bibitem[{Neakrase} {et~al.}(2016){Neakrase}, {Balme}, {Esposito},
  {Kelling}, {Klose}, {Kok}, {Marticorena}, {Merrison}, {Patel}, and
  {Wurm}]{Neakrase2016}
{Neakrase}, L.D.V.; {Balme}, M.R.; {Esposito}, F.; {Kelling}, T.; {Klose}, M.;
  {Kok}, J.F.; {Marticorena},~B.; {Merrison},~J.; {Patel}, M.; {Wurm}, G.
\newblock {Particle Lifting Processes in Dust Devils}.
\newblock {\em Space Sci. Rev.} {\bf 2016}, {\em 203},~347--376,
\newblock
  doi:{\changeurlcolor{black}\href{https://doi.org/10.1007/s11214-016-0296-6}{\detokenize{10.1007/s11214-016-0296-6}}}.

\bibitem[{Nuth} {et~al.}(1994){Nuth}, {Berg}, {Faris}, and
  {Wasilewski}]{Nuth1994}
{Nuth}, J.A.; {Berg}, O.; {Faris}, J.; {Wasilewski}, P.
\newblock {Magnetically enhanced coagulation of very small iron grains}.
\newblock {\em Icarus} {\bf 1994}, {\em 107},~155,
\newblock
  doi:{\changeurlcolor{black}\href{https://doi.org/10.1006/icar.1994.1013}{\detokenize{10.1006/icar.1994.1013}}}.

\bibitem[{N{\"u}bold} {et~al.}(2003){N{\"u}bold}, {Poppe}, {Rost},
  {Dominik}, and {Glassmeier}]{Nuebold2003}
{N{\"u}bold}, H.; {Poppe}, T.; {Rost}, M.; {Dominik}, C.; {Glassmeier}, K.H.
\newblock {Magnetic aggregation. II. Laboratory and microgravity experiments}.
\newblock {\em Icarus} {\bf 2003}, {\em 165},~195--214,
\newblock
  doi:{\changeurlcolor{black}\href{https://doi.org/10.1016/S0019-1035(03)00153-2}{\detokenize{10.1016/S0019-1035(03)00153-2}}}.

\bibitem[{Dominik} and {N{\"u}bold}(2002)]{Dominik2002}
{Dominik}, C.; {N{\"u}bold}, H.
\newblock {Magnetic Aggregation: Dynamics and Numerical Modeling}.
\newblock {\em Icarus} {\bf 2002}, {\em 157},~173--186,
\newblock
  doi:{\changeurlcolor{black}\href{https://doi.org/10.1006/icar.2002.6813}{\detokenize{10.1006/icar.2002.6813}}}.

\bibitem[{Dudorov} and {Khaibrakhmanov}(2014)]{Dudorov2014}
{Dudorov}, A.E.; {Khaibrakhmanov}, S.A.
\newblock {Fossil magnetic field of accretion disks of young stars}.
\newblock {\em \mbox{Astrophys. Space Sci.}} {\bf 2014}, {\em 352},~103--121,
\newblock
  doi:{\changeurlcolor{black}\href{https://doi.org/10.1007/s10509-014-1900-4}{\detokenize{10.1007/s10509-014-1900-4}}}.

\bibitem[{Brauer} {et~al.}(2017){Brauer}, {Wolf}, and {Flock}]{Brauer2017}
{Brauer}, R.; {Wolf}, S.; {Flock}, M.
\newblock {Magnetic fields in circumstellar disks. The potential of Zeeman
  observations}.
\newblock {\em Astron. Astrophys.} {\bf 2017}, {\em 607},~A104,
\newblock
  doi:{\changeurlcolor{black}\href{https://doi.org/10.1051/0004-6361/201731140}{\detokenize{10.1051/0004-6361/201731140}}}.

\bibitem[{Hubbard}(2014)]{Hubbard2014}
{Hubbard}, A.
\newblock {Explaining Mercury's density through magnetic erosion}.
\newblock {\em Icarus} {\bf 2014}, {\em 241},~329--335,
\newblock
  doi:{\changeurlcolor{black}\href{https://doi.org/10.1016/j.Icarus.2014.06.032}{\detokenize{10.1016/j.Icarus.2014.06.032}}}.

\bibitem[{Kruss} and {Wurm}(2018)]{Kruss2018}
{Kruss}, M.; {Wurm}, G.
\newblock {Seeding the formation of Mercurys: An iron sensitive bouncing
  barrier in disk magnetic fields}.
\newblock {\em Icarus} {\bf 2018}, submitted.

\end{thebibliography}

\end{document}